\documentclass[12pt]{article}
\usepackage{amssymb}
\usepackage{epsfig}
\usepackage{float}
\usepackage{graphicx}
\usepackage{amsmath}
\newcommand{\nua}[1]{\ensuremath{\rlap
           {\kern-2.5pt\ensuremath
           {\overset{\scriptscriptstyle(-)}{\phantom{\nu}}}}
           {\ensuremath{{\nu}_{#1}}}}}

\begin{document}

\begin{center}
{\bf Neutrino oscillations: brief history and present status}
\end{center}

\vspace{.7cm}

\begin{center}
{ \bf Samoil M. Bilenky}
\vspace{.1cm}

 {\em  Joint Institute for Nuclear
Research, Dubna, R-141980, Russia\\}
\end{center}
\begin{abstract}
A brief  review of the problem of neutrino masses and oscillations is given. In the beginning we present an early history of neutrino masses, mixing and oscillations. Then we discuss all  possibilities of neutrino masses and mixing (neutrino mass terms). The phenomenology of neutrino oscillations in vacuum is considered in some details. We present also the neutrino oscillation data and the seesaw mechanism of the neutrino mass generation.
\end{abstract}

\section{Introduction}
Discovery of neutrino oscillations in the atmospheric Super-Kamiokande \cite{SK}, solar SNO \cite{SNO}, reactor KamLAND \cite{Kamland}, and other neutrino experiments \cite{Davis,Gallex,Sage}  opened a new
era in neutrino physics: era of investigations of neutrino properties (masses, mixing, nature, etc).

First ideas of neutrino masses, mixing and oscillations belong to Bruno Pontecorvo. He came to an idea of neutrino oscillations in 1957-58 \cite{BPonte57,BPonte58} soon after parity violations in the weak interaction was discovered and the two-component neutrino theory was proposed by Landau \cite{Landau}, Lee and Yang\cite{LeeYang} and Salam \cite{Salam}.

B. Pontecorvo was impressed by a  possibility of
$K^{0}\rightleftarrows\bar K^{0}$ oscillations suggested by
Gell-Mann and Pais \cite{GMPais}. This possibility  was based on the following facts:
\begin{enumerate}
\item The strangeness is conserved in the strong interaction,  $K^{0}$ and $\bar K^{0}$ , eigenstates of the Hamiltonian of the strong interaction, are particles with strangeness +1 and -1, respectively.
\item Weak interaction, in which strangeness is not conserved, induce transitions between $K^{0}$ and $\bar K^{0}$.
\end{enumerate}
As a result of 1. and 2. (assuming that $CP$ is conserved) for the states of particles with definite masses and widths, eigenstates of the total Hamiltonian,  we have
\begin{equation}\label{K12}
| K_{1}^{0}\rangle =\frac{1}{\sqrt{2}}(|K^{0}\rangle + |\bar K^{0}\rangle)\quad | K_{2}^{0}\rangle =\frac{1}{\sqrt{2}}(|K^{0}\rangle - |\bar K^{0}\rangle)
\end{equation}
If in a strong interaction process  $K^{0}$ ($\bar K^{0}$'s) with definite momentum is produced, after the time $t$
we have
\begin{equation}\label{1K12}
| K^{0}\rangle_{t} = \frac{1}{\sqrt{2}}(e^{-i\lambda_{1}t}|K_{1}^{0}\rangle + e^{-i\lambda_{2}t}|\bar K_{2}^{0}\rangle) =
g_{+}(t)|K^{0}\rangle+g_{-}(t)|\bar K^{0}\rangle
\end{equation}
and
\begin{equation}\label{2K12}
|\bar K^{0}\rangle_{t} = \frac{1}{\sqrt{2}}(e^{-i\lambda_{1}t}|K_{1}^{0}\rangle - e^{-i\lambda_{2}t}|\bar K_{2}^{0}\rangle) =
g_{-}(t)|K^{0}\rangle+g_{+}(t)|\bar K^{0}\rangle.
\end{equation}
Here
\begin{equation}\label{3K12}
g_{\pm}(t)=\frac{1}{2}(e^{-i\lambda_{1}t}\pm e^{-i\lambda_{2}t}),
\end{equation}
where $\lambda_{1,2}=m_{1,2}-i\frac{1}{2}\Gamma_{1,2}$, $m_{1,2}$ and $\Gamma_{1,2}$ are masses and total widths of $ K_{1,2}^{0}$. Formulas (\ref{2K12}) and (\ref{3K12}) describe periodical transitions (oscillations) $K^{0}\rightleftarrows\bar K^{0}$.

In the  paper \cite{BPonte57} B. Pontecorvo  raised the following question:
 ''...are there exist other (except $K^{0}$ and $\bar K^{0}$) "mixed" neutral particles (not necessarily elementary ones) which are not identical to their corresponding antiparticles and for which particle $\leftrightarrows$ antiparticle transitions are not strictly forbidden".
He was interested in leptons and came to a conclusion that such mixed (non elementary)  particles could be  muonium $(\mu^{+}-e^{-})$ and antimuonium $(\mu^{-}-e^{+})$. At that time it was not known that  exist different types of neutrinos (the fact that exist $\nu_{e}$ and $\nu_{\mu}$ was discovered in 1962 in the  Brookhaven neutrino experiment \cite{Brookhaven}). In the case of one neutrino in the second order of the perturbation theory over $G_{F}$
transitions
\begin{equation}\label{muonium}
 (\mu^{+}-e^{-})\to \nu+\bar\nu\to (\mu^{-}-e^{+}).
 \end{equation}
are allowed. The paper \cite{BPonte57} was dedicated to the consideration of  muonium $\rightleftarrows$ antimuonium transitions. In the same paper B. Pontecorvo  mentioned a possibility of neutrino oscillations:  ``If the theory of the two-component neutrino is not valid (which is hardly probable at present)
and if the conservation law for the neutrino charge took no place, neutrino $\to $ antineutrino transitions in vacuum
would be in principle possible."

The two-component neutrino theory, was successfully confirmed in 1958 by the experiment \cite{GGS} on the measurement of neutrino helicity. According to this theory for one type of neutrino exist only left-handed neutrino $\nu_{L}$
and right-handed antineutrino $\bar\nu_{R}$. Transitions between these states are forbidden by the conservation of the total angular momentum. Nevertheless {\em in 1958 B. Pontecorvo published the first paper on neutrino oscillations \cite{BPonte58}.} The reason for that paper was some rumor which reached B. Pontecorvo at that time. In 1957-58  R. Davis performed a reactor experiment\cite{Davis} on the search  for production of $^{37}\rm{Ar}$ in the process
\begin{equation}\label{DavisR}
 \mathrm{"reactor ~antineutrino"}+^{37}\rm{Cl} \to e^{-}+^{37}\rm{Ar}
\end{equation}
in which the lepton number is violated. The rumor reached  B.Pontecorvo that Davis had observed
such events. Pontecorvo, who was thinking about a possibility of neutrino oscillations, suggested that these "events" could be due to transitions $\bar\nu_{R}\to \nu_{R}$. This suggestion contradicted to the two-component neutrino theory. In fact, according to this theory only the field $\nu_{L}(x)$ enters in the weak interaction Lagrangian. Thus, from the point of view of the two-component theory $\nu_{R}$ and  $\bar\nu_{L}$ were noninteracting "sterile" particles. However, in order to explain Davis "events"  Pontecorvo had to assume that "a definite fraction of particles can induce the reaction (\ref{DavisR})". He assumed (in analogy with (\ref{K12})) that states $\bar\nu_{R}$ and $\nu_{R}$ states are mixed:
\begin{equation}\label{numix}
|\bar\nu_{R}\rangle =\frac{1}{\sqrt{2}}(|\nu_{1}\rangle + |\nu_{2}\rangle)\quad
|\nu_{R}\rangle =\frac{1}{\sqrt{2}}(|\nu_{1}\rangle - |\nu_{2}\rangle),
\end{equation}
where $|\nu_{1}\rangle$ and $|\nu_{2}\rangle$ are states of neutrinos with masses $m_{1}$ and $m_{2}$. B. Pontecorvo wrote in the paper  \cite{BPonte58}: :``...neutrino may be a particle mixture and consequently there is a possibility of real transitions neutrino $\to$ antineutrino in vacuum, provided that the lepton (neutrino) charge is not conserved. This means that the
neutrino and antineutrino are {\em mixed} particles, i.e., a symmetric and
antisymmetric combination of two truly neutral Majorana particles $\nu_1$
and $\nu_2$...this possibility became of some interest in connection with new investigations of inverse $\beta$-processes."

As it is well known, (anti)neutrino was discovered in the fifties in the Reines and Cowan experiments \cite{Reines}
in which  the reaction
\begin{equation}\label{ReinesC}
\bar\nu+p \to e^{+}+n.
\end{equation}
with reactor antineutrinos was observed.

In the paper \cite{BPonte58}  B. Pontecorvo  considered disappearance of reactor antineutrinos due to neutrino oscillations. He wrote: ``...the cross section of the process $\bar\nu +p\to e^{+}+n$ with
$\bar\nu$ from reactor must be smaller than expected. This is due to the fact that the neutral lepton beam which at the source is capable of inducing the reaction  changes its composition on the way from the reactor to the detector."

Starting from the  first paper on the neutrino oscillations B. Pontecorvo believed that neutrinos have small masses  and oscillate. He wrote in \cite{BPonte58}: ``Effects of transformation of neutrino into antineutrino and vice versa may be unobservable in the laboratory  because of the large values of R (oscillation length) but will certainly occur, at least, on an astronomical scale."

At a later stage of the Davis experiment  the anomalous "events" (\ref{DavisR}) disappeared and only an upper bound of the cross section of the reaction (\ref{DavisR}) was obtained in \cite{Davis}. Continuing to think about neutrino oscillations, B. Pontecorvo understood that
 $\nu_{R}$ and  $\bar\nu_{L}$ are noninteracting, sterile particles and that he considered in  \cite{BPonte58} oscillations between active antineutrino $\bar\nu_{R}$ and sterile neutrino $\nu_{R}$.\footnote{ Such oscillations are very modern nowadays. They are considering as a possible explanation of the the  "reactor neutrino anomaly" \cite{Ranomaly}} The terminology "sterile neutrino", which is standard nowadays, was introduced by B. Pontecorvo in {\em the next paper on neutrino oscillations \cite{BPonte67} which was published  in 1967.}

 At that time the phenomenological V-A theory  \cite{MarSud,FeyGel}  was well established, $K^{0}\rightleftarrows \bar K^{0}$ oscillations had been observed and it was proved that (at least) two types of neutrinos $\nu_{e}$ and $\nu_{\mu}$ existed in nature. Pontecorvo wrote: ``If the lepton charge is not an exactly conserved quantum number, and the neutrino mass is different from zero, oscillations similar to those in $K^{0}$ beams become possible in neutrino beams". He discussed all possible transitions between two types of neutrinos,   $\nu_{e}\rightleftarrows\bar\nu_{e L}$, $\nu_{\mu}\rightleftarrows\bar\nu_{\mu L}$ etc. which transform ``active particles into particles, which from the point of view of ordinary weak processes, are sterile"  and transitions between active neutrinos $\nu_{\mu}\rightleftarrows\nu_{e}$. He pointed out that in this last case not only the disappearance of  $\nu_{\mu}$  but also the appearance of $\nu_{e}$ can be observed.

In the 1967 paper \cite{BPonte67} B.Pontecorvo  discussed the effect of neutrino oscillations for the solar neutrinos.
``From an observational point of view the ideal object is the sun. If
the oscillation length is smaller than the radius of the sun region
effectively producing neutrinos, direct
oscillations will be smeared out and unobservable. The only effect on
the earth's surface would be that the flux of observable sun neutrinos
must be two times smaller than the total (active and sterile) neutrino flux."
{\em Thus, Bruno Pontecorvo anticipated the solar neutrino problem.}
When three years later in 1970 the first results of the Davis experiment on the detection of the solar neutrinos were obtained \cite{Davis} it occurred that the detected flux of  solar neutrinos was about 2-3 times smaller than the flux predicted by the Standard solar model. It was  soon commonly accepted that among different possible astrophysical (and particle physics) explanations of the problem neutrino oscillations  was the most natural one (see \cite{BilPonteCom}).

{\em Next paper on the neutrino oscillation was published in 1969 by V. Gribov and B.Pontecorvo }\cite{GPonte69}.
They considered a scheme of neutrino mixing and oscillations with four neutrino and antineutrino states:  two left-handed states of neutrinos $\nu_{e}$ and $\nu_{\mu}$  and two  right-handed states of antineutrinos
$\bar\nu_{e}$ and $\bar\nu_{\mu}$. The main assumption of  Gribov and Pontecorvo was that there are no sterile neutrino states.

It was assumed in \cite{GPonte69} that in addition to the standard charged current $V-A$ interaction with the lepton current
\begin{equation}\label{CC}
j_{\alpha}=2(\bar\nu_{eL}\gamma_{\alpha}e_{L}+\bar\nu_{\mu L}\gamma_{\alpha}\mu_{L})
\end{equation}
in the total Lagrangian enters {\em an effective Lagrangian of an interaction between neutrinos which violates electron $L_{e}$ and muon $L_{\mu}$ lepton numbers} (in modern terminology neutrino mass term).
 After the diagonalization of the most general effective Lagrangian of  such type the following mixing relations were found
\begin{equation}\label{GPonte}
 \nu_{eL}(x)=\cos\theta \chi_{1L}(x)+\sin\theta \chi_{2L}(x);\quad
\nu_{\mu L}(x)=-\sin\theta \chi_{1L}(x)+\cos\theta \chi_{2L}(x).
\end{equation}
Here $\chi_{1,2}(x)$ are fields of the Majorana neutrinos with masses $m_{1,2}$ and $\theta$ is the mixing angle. All these parameters are determined by the parameters which characterize the effective Lagrangian.

The authors obtained the following expression for the $\nu_{e}\to \nu_{e}$ survival probability in vacuum (in modern notations)
\begin{equation}\label{survival}
P(\nu_{e}\to \nu_{e})=1-\frac{1}{2}\sin^{2}2\theta~(1-\cos\frac{\Delta m^{2}L}{2E})
\end{equation}
($\Delta m^{2}=|m^{2}_{2}-m^{2}_{1}|$) and  applied the  developed formalism to  solar neutrinos.
They considered the possibility of the maximal mixing $\theta=\frac{\pi}{4}$  as the most simple
and attractive one. In this case the averaged observed flux of solar neutrinos is equal to 1/2 of the predicted flux.

 Bruno Pontecorvo and my work  on neutrino masses, mixing and  oscillations started in 1975. The first paper \cite{BilPonte76} was based on the idea of {\em quark-lepton analogy}.
It had been established at that time that the charged current of quarks has the form (the case of four quarks)
\begin{equation}\label{CCq}
j^{CC}_{\alpha}(x)=2[\bar u_{L}(x)\gamma_{\alpha}d^{c}_{L}(x)+\bar c_{L}(x)\gamma_{\alpha}s^{c}_{L}(x)],
\end{equation}
where
\begin{equation}\label{Mixq}
d^{c}_{L}(x)=\cos\theta_{C}d_{L}(x)+\sin\theta_{C}s_{L}(x),
\quad
s^{c}_{L}(x)=-\sin\theta_{C}d_{L}(x)+\cos\theta_{C}s_{L}(x)
\end{equation}
are Cabibbo-GIM mixtures of $d$ and $s$ quarks  and $\theta_{C}$ is the Cabibbo angle.

The lepton charged current (\ref{CC}) has the same form as the quark charged current (same coefficients, left-handed components of the fields). In order to make the analogy between quarks and leptons
complete it was natural to assume that current neutrino fields $\nu_{e L}(x)$ and
$\nu_{\mu L}(x)$ are also mixed:
\begin{equation}\label{Mixl}
\nu_{eL}(x)=\cos\theta \nu_{1L}(x)+\sin\theta \nu_{2L}(x),\quad
\nu_{\mu L}(x)=-\sin\theta \nu_{1L}(x)+\cos\theta \nu_{2L}(x).
\end{equation}
Here $\theta$ is the leptonic mixing angle and $\nu_{1}(x)$ and $\nu_{2}(x)$  (like quark fields)  are {\em Dirac fields} of neutrinos with masses $m_{1}$ and $m_{2}$ .
 This means that the total lepton number $L=L_{e}+L_{\mu}$ is conserved and the neutrinos with definite masses $\nu_{1,2}$  differ from the corresponding antineutrinos $\bar\nu_{1,2}$ by the lepton number ($L(\nu_{1,2})=-L(\bar\nu_{1,2})=1$).

We wrote in \cite{BilPonte76}: ``...in our scheme $\nu_{1}$ and
$\nu_{2}$ are just as leptons and quarks (which, may be, is an attractive feature) while in the Gribov-Pontecorvo scheme \cite{GPonte69} the two neutrinos have a special position among the other fundamental particles".

In 1975 after the  success of the two-component theory  there was still a general belief than neutrinos are massless particles. It is obvious that in this case the mixing (\ref{Mixl}) has no physical meaning.

Our main arguments for neutrino masses were at that time the following
\begin{enumerate}
  \item There is no principle (like gauge invariance in the case of the $\gamma$-quanta) which requires that  neutrino masses must be equal  to zero.
  \item In the framework of the two-component neutrino theory the zero masses of the neutrinos was considered as an argument in favor of the left-handed neutrino fields.  It occurred, however, that in the weak Hamiltonian enter left-handed components of {\em all fields} (the $V-A$ theory). It was more natural after the $V-A$ theory was established to consider neutrinos not as a special massless particles but as  particles with some small masses.
\end{enumerate}
We discussed in \cite{BilPonte76} a  possible value of the mixing angle $\theta$. We argued that
\begin{itemize}
  \item there is no reason for the lepton and Cabibbo mixing angles to be the same.
  \item ``it seems to us that the special values of the mixing angles $\theta=0$ and $\theta=\frac{\pi}{4}$ (maximum mixing) are of the greatest interest."
\end{itemize}
The probabilities of  transitions $\nu_{l}\to \nu_{l'}$ ($l,l'=e,\mu$) are the same in the scheme  of the mixing of  two Majorana neutrinos \cite{GPonte69} and in the scheme of the mixing of  two Dirac neutrinos \cite{BilPonte76}.

In the next paper  \cite{BilPonteNC} we considered the most general neutrino mixing which included not only active
 $\nu_{eL}(x)$ and $\nu_{\mu L}(x)$ fields but also sterile  $\nu_{eR}(x)$ and $\nu_{\mu R}(x)$ neutrino fields (Dirac and Majorana mass term). We applied the developed formalism to solar neutrinos and showed that in this case the maximal suppression of the solar neutrino flux can be $\frac{1}{4}$ (for the two neutrino types).

Ideas of of neutrino masses, mixing and oscillations we discussed before were based on analogy between the weak interaction of lepton and hadrons (later quarks) and on the attractiveness of a minimal scheme of neutrino mixing (no sterile fields, Majorana mass term). {\em In 1962 Maki, Nakagawa and Sakata \cite{MNS}  introduced the neutrino
mixing in the framework of the Nagoya model in which the proton, neutron and $\Lambda$
were considered as bound states of neutrino, electron and muon and a
vector boson $B^{+}$, ``a new sort of matter"}. The authors wrote: ``We assume that there exists a representation
which defines the true neutrinos $\nu_{1}$ and $\nu_{2}$ through
orthogonal transformation
\begin{eqnarray}\label{MNS}
\nu_{1}&=&\cos\delta\nu_{e} - \sin\delta \nu_{\mu}\nonumber\\
\nu_{2}&=&\sin\delta\nu_{e} + \cos\delta \nu_{\mu},
\end{eqnarray}
where $\delta$ is a real constant".

The authors wrote: "Weak
neutrinos must be redefined by a relation
\begin{eqnarray}\label{MNS1}
\nu_{e}&=&\cos\delta\nu_{1} + \sin\delta \nu_{2}\nonumber\\
\nu_{\mu}&=&-\sin\delta\nu_{1} + \cos\delta \nu_{2}.
\end{eqnarray}
An further in the paper "...weak neutrinos are not stable due to the occurrence of virtual transition
$\nu_{e}\rightleftarrows \nu_{\mu}$..."  In connection with the Brookhaven
neutrino experiment \cite{Brookhaven}, which was going on at that time, the authors stressed that : ``... a chain of reactions such as
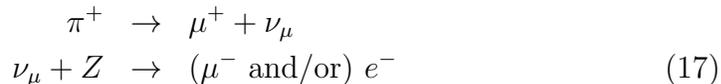
\begin{eqnarray}\label{MNS2}
\pi^{+}&\to &\mu^{+}+\nu_{\mu}\nonumber\\
\nu_{\mu}+Z&\to & (\mu^{-}~\mathrm{and/or})~e^{-}
\end{eqnarray}
is useful to check the two-neutrino hypothesis only when
\begin{equation}\label{MNS3}
   | m_{\nu_{2}}- m_{\nu_{1}}|\leq ~10^{-6}\mathrm{MeV}
\end{equation}
under the conventional geometry of the experiments. Conversely, the
absence of $e^{-}$ will be able not only to verify the two-neutrino
hypothesis but also to provide an upper limit of the mass of the
second neutrino $\nu_{2}$ if the present scheme should be accepted."

In conclusion we mention also  Fritzsch and Minkowski \cite{FM76}
and Eliezer and Swift \cite{ES76}  earlier papers on  the vacuum neutrino oscillations and very important Wolfenstein \cite{Wolf} and Mikheev and Smirnov \cite{MS} papers
in which importance of matter effect for solar and other neutrinos was discovered.

\section{Neutrino mass terms}
In reviews \cite{BilPonte77,BilPetRMP} all possible
schemes of neutrino mixing were considered. The scheme of neutrino
mixing is determined by {\em a neutrino mass term.} There are three
possible neutrino mass terms.
\begin{center}
{\bf Dirac mass term}
\end{center}
Let us assume that
in the total Lagrangian a Dirac mass term enters.
The Dirac mass term is a Lorenz-invariant product of left-handed and
right-handed fields. We have
\begin{equation}\label{Dmst}
 \mathcal{L}^{\mathrm{D}}(x)=-\bar
\nu'_{L}(x)\,M^{\mathrm{D}}\, \nu'_{R}(x)+\mathrm{h.c.}.
\end{equation}
Here
\begin{eqnarray}\label{Dir1}
\nu'_{L}=\left(
\begin{array}{c}
  \nu_{e L} \\
  \nu_{\mu L}\\
  \nu_{\tau L}
\end{array}
\right),\quad \nu'_{R}=\left(
\begin{array}{c}
  \nu_{e R} \\
  \nu_{\mu R}\\
  \nu_{\tau R}
\end{array}
\right),
\end{eqnarray}
 and  $M^{\mathrm{D}}$ is a complex $3\times3$ matrix.

An arbitrary nonsingular matrix $M$ can be diagonalized by a
biunitary transformation
\begin{equation}\label{biunit}
M=U^{\dag}~m~V.
\end{equation}
Here $U$ and $V$ are unitary matrices and $m$ is a diagonal matrix with positive diagonal elements ($m_{ik}=m_{i}\delta_{ik}, m_{i}>0$).

From (\ref{Dmst}) and (\ref{biunit}) for the neutrino mass term we
find
\begin{equation}\label{Dmst1}
 \mathcal{L}^{\mathrm{D}}(x)=- \bar\nu_{ L}(x)\,
m\,\nu_{ R}(x) +\rm{h.c.}=-\bar\nu(x)\,
m\,\nu(x)=-\sum_{i}m_{i}~\bar\nu_{i}(x)\, \nu_{i}(x).
\end{equation}
Here
\begin{eqnarray}
\nu_{L}=U^{\dag}\nu'_{L}=\left(\begin{array}{c}
\nu_{1L} \\
\nu_{2 L}  \\
\nu_{3 L}
\end{array}\right),~~
\nu_{R}=V^{\dag}\nu'_{R}=\left(\begin{array}{c}
\nu_{1R} \\
\nu_{2 R}  \\
\nu_{3 R}
\end{array}\right),~~\nu=\nu_{L}+\nu_{R}=\left(\begin{array}{c}
\nu_{1} \\
\nu_{2 }  \\
\nu_{3 }
\end{array}\right)
\label{Diracmass1}
\end{eqnarray}
After the diagonalization of the matrix $M^{\mathrm{D}}$ we come to
the standard expression for the Dirac neutrino mass term. From
(\ref{Dmst1}) and (\ref{Diracmass1}) it follows that $\nu_{i}(x)$ is
the field of neutrinos with  mass $m_{i}$ and flavor neutrino fields
$\nu_{lL}(x)$ ($l=e,\mu,\tau$) are "mixtures" of the left-handed components of the fields of
neutrinos with definite masses:
\begin{equation}\label{mixing}
  \nu_{lL}(x)=\sum^{3}_{i=1}U_{li}\nu_{iL}(x).
\end{equation}
The fields $\nu_{i}(x)$ are complex (nonhermitian) fields which satisfy the Dirac equation. There are
no additional constraints on them. Thus, the neutrino mass term,  the lepton charged current
\begin{equation}\label{lepCC}
j_{\alpha}=2\sum_{l=e,\mu,\tau}\bar\nu_{lL}\gamma_{\alpha}l_{L}
\end{equation}
and other terms of  total Lagrangian are invariant under
the following {\em phase transformation}
\begin{equation}\label{phaseinv}
  \nu_{i}(x)\to e^{i\Lambda}\nu_{i}(x),\quad l(x)\to e^{i\Lambda}l(x),
\end{equation}
where $\Lambda$ is an arbitrary constant.

From invariance under the  gauge transformation
(\ref{phaseinv}) follows that the total lepton number $L$, the same
for all charged leptons $e,\mu,\tau$ and all neutrinos $\nu_{1},\nu_{2},\nu_{3}$ is
conserved and that neutrinos  $\nu_{i}$  and antineutrinos
$\bar\nu_{i}$ ($i=1,2,3$) are different particles: they differ by the values of
the conserved lepton number. We can choose
\begin{equation}\label{phaseinv1}
  L(\nu_{i})=-L(\bar\nu_{i})=1,\quad  L(l^{-})=-L(l^{+})=1.
\end{equation}
Thus, in the case of the mass term (\ref{Dmst}), {\em neutrinos with
definite masses are Dirac particles.}

Summarizing, we stress the following
\begin{enumerate}
  \item In order to build the Dirac mass term we need the flavor left-handed fields
$\nu_{lL}$ and the (sterile) right-handed fields $\nu_{lR}$.
  \item The fields $\nu_{lL}$ and $\nu_{lR}$ are connected, correspondingly,
 with $\nu_{iL}$ and $\nu_{iR}$ by (different) unitary transformations.
  \item Transitions in vacuum of active to sterile neutrinos
($\nu_{lL}\to \bar\nu_{l'L}$) are forbidden by the conservation of
the total lepton number $L$.
\end{enumerate}

\begin{center}
{\bf Majorana mass term}
\end{center}
The fields $\nu_{L,R}$ are determined by the conditions
\begin{equation}\label{L,R}
\gamma_{5}\nu_{L,R}=\mp \nu_{L,R}.
\end{equation}
Let us consider
\begin{equation}\label{Cconj}
(\nu_{L,R})^{c}=C\bar\nu_{L,R}^{T},
\end{equation}
where $C$ is the matrix of the charge
conjugation which satisfies the relations
\begin{equation}\label{Cconj1}
C\gamma^{T}_{\alpha}C^{-1}=-\gamma_{\alpha},\quad C^{T}=-C
\end{equation}
We have
\begin{equation}\label{Cconj2}
\gamma_{5}(\nu_{L,R})^{c}=C(\bar\nu_{L,R}\gamma_{5})^{T} =-
C\overline{(\gamma_{5}\nu_{L,R})}^{T}=\pm (\nu_{L,R})^{c}.
\end{equation}
Thus, $(\nu_{L,R})^{c}$ is right-handed (left-handed) field.

The neutrino mass term (a product of left-handed and right-handed
fields) can have the following form
\begin{equation}\label{Mjmst}
 \mathcal{L}^{\mathrm{M}}(x)=-\frac{1}{2} \bar\nu'_{ L}(x)\,
M^{\mathrm{M}}\,(\nu'_{ L}(x))^{c} +\rm{h.c.},
\end{equation}
where $M^{\mathrm{M}}$ is a complex nondiagonal  matrix. It is easy to see  that $M^{\mathrm{M}}$ is a symmetrical matrix.\footnote{Taking into account
the Fermi-Dirac statistics of the fermion fields $\nu'_{L}$ we have
$$\bar\nu'_{L} M^{\mathrm{M}}(\nu'_{L})^{c}=-\bar\nu'_{
L}C^{T}(M^{\mathrm{M}})^{T}(\bar\nu'_{L})^{T}=\bar\nu'_{
L}(M^{\mathrm{M}})^{T}C(\bar\nu'_{L})^{T},~\mathrm{i.e.}~M^{\mathrm{M}}=(M^{\mathrm{M}})^{T}.$$}

A complex symmetrical matrix can be diagonalized with the help of {\em one
unitary matrix}. We have
\begin{equation}\label{Mjmst3}
M^{\mathrm{M}}=U~m~U^{T},
\end{equation}
where $U^{\dag}U=1$ and $m_{ik}=m_{i}\delta_{ik},~~m_{i}>0$.

From (\ref{Mjmst}) and (\ref{Mjmst3}) for the neutrino mass term we
obtain the following expression
\begin{equation}\label{Mjmst4}
 \mathcal{L}^{\mathrm{M}}=-\frac{1}{2}\overline{(U^{\dag}\nu'_{
L})}~m~(U^{\dag}\nu'_{
L})^{c}+\rm{h.c.}=-\frac{1}{2}\bar\nu^{m}~m~\nu^{m}.
\end{equation}
Here
\begin{eqnarray}
\nu^{m}=U^{\dag}\nu'_{L}+(U^{\dag}\nu'_{L})^{c}=
\left(\begin{array}{c}
\nu_{1} \\
\nu_{2 }  \\
\nu_{3 }.
\end{array}\right). \label{Mjmst5}
\end{eqnarray}
From (\ref{Mjmst4}) and (\ref{Mjmst5}) it follows that
\begin{equation}\label{Mjmst6}
\mathcal{L}^{\mathrm{M}}=-\frac{1}{2}\sum^{3}_{i=1}m_{i}~\bar\nu_{i}~\nu_{i}.
\end{equation}
and the field $\nu_{i}(x)$ satisfies the Majorana
condition
\begin{equation}\label{Mjmst7}
\nu_{i}(x)= \nu^{c}_{i}(x)=C\bar\nu^{T}_{i}(x)
\end{equation}
Thus, $\nu_{i}(x)$ is  the field of the Majorana neutrino with the mass $m_{i}$.

From (\ref{Mjmst5}) we conclude that  flavor
fields $\nu_{lL}(x)$ are mixtures of left-handed components of
Majorana fields with definite masses
\begin{equation}\label{Mjmst8}
\nu_{lL}(x)=\sum^{3}_{i=1}U_{li}\nu_{iL}(x).
\end{equation}
If neutrino field satisfies the Majorana condition
(\ref{Mjmst7}) in this case there is no notion of neutrino and
antineutrino: neutrino and antineutrino are identical ($\nu_{i}\equiv\bar\nu_{i}$).
It is obvious that  we came to the fields of Majorana neutrinos with
definite masses because in the case of the neutrino mass term  (\ref{Mjmst})
 there
is no invariance under the global gauge transformations
\begin{equation}\label{phaseinv1}
  \nu_{lL}(x)\to e^{i\Lambda}\nu_{lL}(x),\quad l(x)\to e^{i\Lambda}l(x).
\end{equation}
and, consequently, no conserved lepton number which could distinguish neutrino and antineutrino.

The mass term (\ref{Mjmst})
provides {\em the minimal scheme of neutrino mixing}: only flavor
neutrino fields $\nu_{lL}(x)$ enter into the Lagrangian.\footnote{There are no right-handed sterile neutrino fields in the Lagrangian: point of view advocated by Gribov and Pontecorvo\cite{GPonte69}.} We will see later
that the Majorana mass term is generated by the standard seesaw
mechanism \cite{seesaw} of the generation of small neutrino
masses.

\begin{center}
{\bf Dirac and Majorana mass term}
\end{center}
If we assume that in the  mass term enter flavor left-handed fields
$\nu_{lL}(x)$, right-handed sterile fields $\nu_{lR}(x)$ and the lepton
number $L$ is not conserved we come to the most general Dirac and Majorana mass neutrino mass
term \cite{BilPonteNC}
\begin{equation}\label{DMjmst}
\mathcal{L}^{\mathrm{D+M}}= -\frac{1}{2}\,
\bar\nu'_{L}\,M^{\mathrm{M}}_{L} ( \nu'_{L})^{c}- \bar
\nu'_{L}\,M^{\mathrm{D}}\, \nu'_{R}
-\frac{1}{2}\,\overline{(\nu'_{R})^{c}}\,M^{\mathrm{M}}_{R} \nu'_{R}
+\mathrm{h.c.},
\end{equation}
where $M^{\mathrm{M}}_{L}$, $M^{\mathrm{M}}_{R}$  are complex, non-diagonal, symmetrical 3 $\times$ 3 matrices, $M^{\mathrm{D}}$
is a complex non-diagonal 3 $\times$ 3 matrix and  columns
$\nu'_{L}$ and $\nu'_{R}$ are given by (\ref{Dir1}). The first term of (\ref{DMjmst}) is the left-handed Majorana mass
term, the second term is the Dirac mass term and the third term is
the right-handed Majorana mass term. If the mass
term (\ref{DMjmst}) enters into the total Lagrangian in this case the
total lepton number $L$ is not conserved.

It is obvious that
\begin{equation}\label{DMjmst2}
\bar \nu'_{L}\,M^{\mathrm{D}}\,
\nu'_{R}=(\nu'_{R})^{T}\,C^{-1}(M^{\mathrm{D}})^{T}\,C
(\bar\nu'_{L})^{T}=\overline{(\nu'_{R})^{c}}(M^{\mathrm{D}})^{T}\nu'_{L})^{c}.
\end{equation}
Taking into account this relation, we can present the Dirac and
Majorana mass term (\ref{DMjmst}) in the form
\begin{equation}\label{DMjmst3}
\mathcal{L}^{\mathrm{D+M}}= -\frac{1}{2}\,
\bar n_{L}\,M^{\mathrm{M+D}} ( n_{L})^{c}+\mathrm{h.c.}.
\end{equation}
Here
\begin{eqnarray}
n_{L}=\left(
\begin{array}{c}
\nu'_{L}\\
(\nu'_{R})^{c}\\
\end{array}
\right).\label{DMjmst4}
\end{eqnarray}
and
\begin{eqnarray}
M^{\mathrm{M+D}}=\left(
  \begin{array}{cc}
  M^{\mathrm{M}}_{L} & M^{\mathrm{D}} \\
   (M^{\mathrm{D}})^{T}&M^{\mathrm{M}}_{R}\\
\end{array}
\right). \label{DMjmst5}
\end{eqnarray}
 is a symmetrical $6\times6$ matrix. We have
\begin{equation}\label{DMjmst6}
M^{\mathrm{M+D}}=U~m~U^{T},
\end{equation}
where $U$ is a unitary $6\times6$ matrix and
$m_{ik}=m_{i}\delta_{ik}$. From (\ref{DMjmst6}) for the Dirac and
Majorana mass term we obtain the following expression
\begin{equation}\label{DMjmst7}
\mathcal{L}^{\mathrm{D+M}}(x)=-\frac{1}{2}\bar\nu^{m}(x)~m~\nu^{m}(x)
=-\frac{1}{2}\sum^{6}_{i=1}m_{i}~\bar\nu_{i}(x)~\nu_{i}(x),
\end{equation}
where
\begin{eqnarray}
\nu^{m}(x)=U^{\dag}n_{L}(x)+(U^{\dag} n_{L}(x))^{c}=
\left(\begin{array}{c}
\nu_{1}(x) \\
\nu_{2 }(x)  \\
.\\
.\\
.\\
\nu_{6 }(x).
\end{array}\right). \label{DMjmst8}
\end{eqnarray}
 It is obvious that the field $\nu_{i}(x)$
satisfies the  condition
\begin{equation}\label{DMjmst9}
\nu_{i}(x)=\nu^{c}_{i}(x)=C\bar \nu^{T}_{i}(x).
\end{equation}
and  is the field of the Majorana neutrino with
mass $m_{i}$.

From (\ref{DMjmst9}) it follows that in the case of the Dirac and
Majorana mass term we have the following generalized neutrino mixing
relations
\begin{equation}\label{DMjmst11}
\nu_{l L}(x)=\sum^{6}_{i=1}U_{l i}\,\nu_{i L}(x),\quad (\nu_{l
R}(x))^{c}= \sum^{6}_{i=1}U_{\bar {l} i}\,\nu_{i L}(x)~~(l=e,\mu,\tau)
\end{equation}
Thus, flavor fields $\nu_{l L}(x)$ are combinations of
left-handed components of  six Majorana  fields. These six components
are connected by the unitary transformations with sterile fields
$(\nu_{l R}(x))^{c}$.

Let us notice that the seesaw mechanism of neutrino mass generation \cite{seesaw}
is based on the Dirac and Majorana mass term. In the seesaw case in
the mass spectrum of the Majorana particles there are three light
neutrino masses $m_{i}$  and three heavy masses $M_{i}\gg
m_{i}$ ( $i=1,2,3$).

If all Majorana masses are small, in this case transitions of flavor
neutrinos $\nu_{e},\nu_{\mu},\nu_{\tau}$ into sterile states
 become possible. Let us notice that at present exist
experimental indications  in favor of such
transitions. These indications will be checked in future experiments (see \cite{sterile}).

\section{Phenomenology of neutrino oscillations in vacuum}
In the case of the neutrino mixing for  the   flavor   neutrino  field $\nu_{l L}(x)$ which enters into the standard
leptonic charged current
\begin{equation}\label{1lepCC}
j_{\alpha}(x)=2\sum_{l=e,\mu,\tau}\nu_{l L}(x)\gamma_{\alpha}l_{L}(x)
\end{equation}
we have
\begin{equation}\label{3numix}
\nu_{l L}(x)=\sum_{i} U_{l i}\,\nu_{i L}(x),
\end{equation}
where $\nu_{i }(x)$ is the field of neutrino (Dirac or Majorana) with mass $m_{i}$. If neutrino mass-squared differences are small the state of the flavor neutrino $\nu_{l}$ produced in CC weak decays and reactions together with $l^{+}$
is given by the following coherent superpositions of the states of neutrinos with definite masses
\begin{equation}\label{flavstate}
|\nu_{l}\rangle=\sum^{3}_{i=1}U^{*}_{li}~|\nu_{i}\rangle.
\end{equation}
Here $|\nu_{i}\rangle$ is the state of neutrino with momentum $\vec{p}$ and energy $E_{i}=\sqrt{p^{2}+m_{i}^{2}}$, eigenstate of the free Hamiltonian
\begin{equation}\label{eigenstate}
H_{0}~|\nu_{i}\rangle=E_{i}~|\nu_{i}\rangle.
\end{equation}
In connection with the relation (\ref{flavstate}), which is a basic relation of the theory of the neutrino oscillations, we will make the following remarks
\begin{enumerate}
\item It is impossible to reveal different neutrino masses in the production (and detection) process if the following condition is satisfied (see \cite{BilFP}) $$L_{ik}\gg d.$$ Here $L_{ik}=4\pi\frac{E}{|\Delta m_{ik}^{2}|}$ ($i\neq k$) is the oscillation length
($\Delta m_{ik}^{2}= m_{k}^{2}-m_{i}^{2}$, $E$ is the neutrino energy) and $d$ is a quantum mechanical size of the neutrino source. This condition follows from  the Heisenberg uncertainty relation and is satisfied in all current neutrino oscillation experiments (for the atmospheric neutrinos $L_{23}\simeq 10^{3}$ km, for the long baseline reactor neutrinos $L_{12}\simeq 10^{2}$ km)
\item Flavor neutrino states  are orthogonal and normalized
$$\langle \nu_{l'}|\nu_{l}\rangle=\delta_{l'l}$$
and do not depend on process in which neutrino is produced
\item Flavor antineutrino states are given by the expression
\begin{equation}\label{flavantinu}
|\bar\nu_{l}\rangle=\sum^{3}_{i=1}U_{li}~|\bar\nu_{i}\rangle.
\end{equation}
\item In the case of the active and sterile neutrinos we have
\begin{equation}\label{flavsterstate}
|\nu_{l}\rangle=\sum^{3+n_{s}}_{i=1}U^{*}_{li}~|\nu_{i}\rangle~~(l=e,\mu,\tau)\quad |\nu_{s}\rangle=\sum^{3+n_{s}}_{i=1}U^{*}_{si}~|\nu_{i}\rangle ~~ s=s_{1},...s_{n_{s}}
\end{equation}
\end{enumerate}
Evolution of states in QFT is  given by the Schrodinger equation
\begin{equation}\label{Schrod}
i\frac{\partial}{\partial t}|\Psi(t)\rangle=H~|\Psi(t)\rangle.
\end{equation}
If at the time $t=0$ flavor neutrino $\nu_{l}$ is produced, at the time $t$ for the neutrino state we have
\begin{equation}\label{Schrod1}
|\nu_{l}\rangle_{t}=e^{-iH_{0}t}
\sum^{3}_{i=1}U^{*}_{li}~|\nu_{i}\rangle=
\sum_{i}|\nu_{i}\rangle e^{-iE_{i}t}~U^{*}_{li}
\end{equation}
Thus, {\em in the case of the neutrino mixing neutrino is described  by a nonstationary state} (by a superposition of states with different energies).

Neutrinos are detected via observation of weak processes $\nu_{l}+N\to l^{-}+X$ etc. We have
\begin{equation}\label{Schrod2}
|\nu_{l}\rangle_{t}=\sum_{l'}|\nu_{l'}\rangle (\sum^{3}_{i=1}U_{l'i}~e^{-iE_{i}t}~U^{*}_{li}).
\end{equation}
From this relation for the  probability of the $\nu_{l}\to\nu_{l'}$  transition during the time $t$ we find the following  expression
\begin{equation}\label{Schrod3}
 P(\nu_{l}\to\nu_{l'})=
|\sum^{3}_{i=1}U_{l'i}~e^{-iE_{i}~t}~U^{*}_{li}|^{2}.
\end{equation}
Analogously for the probability of the $\bar\nu_{l}\to \bar\nu_{l'}$  transition we have
\begin{equation}\label{Schrod4}
 P(\bar\nu_{l}\to \bar\nu_{l'})=
|\sum^{3}_{i=1}U^{*}_{l'i}~e^{-iE_{i}~t}~U_{li}|^{2}.
\end{equation}
In the general case which include not only active flavor neutrinos but also sterile neutrinos we have
\begin{equation}\label{genexp}
  P(\nu_{\alpha}\to \nu_{\alpha'})=|\sum^{3+n_{s}}_{i=1}U_{\alpha'i}~e^{-iE_{i}t}~
U^{*}_{\alpha i}|^{2}.
\end{equation}
and
\begin{equation}\label{genexp1}
  P(\bar\nu_{\alpha}\to \bar\nu_{\alpha'})=|\sum^{3+n_{s}}_{i=1}U^{*}_{\alpha'i}~e^{-iE_{i}t}~
U_{\alpha i}|^{2},
\end{equation}
where indexes $\alpha,\alpha'$ run over $e,\mu,\tau,s_{1},...s_{n_{s}}$. From (\ref{genexp}) and (\ref{genexp1})
we  obtain the following standard expression for the neutrino transition probability
\begin{eqnarray}\label{genexp5}
P(\nua{\alpha}\to \nua{\alpha'})&=&\delta_{\alpha' \alpha }
-2\sum_{i>k}\mathrm{Re}~U_{\alpha' i}U^{*}_{\alpha i}U_{\alpha'
k}^{*}U_{\alpha k}~(1-\cos\frac{\Delta m^{2}_{ki}L}{2E})\nonumber\\
 &\pm&
2\sum_{i>k}\mathrm{Im}~U_{\alpha' i}U^{*}_{\alpha i}U_{\alpha'
k}^{*}U_{\alpha k}~\sin\frac{\Delta m^{2}_{ki}L}{2E}
\end{eqnarray}
 In (\ref{genexp5}) we took into account that for the ultrarelativistic neutrinos $L\simeq t$ ($L$ is the distance between neutrino production and detection points) and $E_{i}-E_{k}\simeq \frac{\Delta m^{2}_{ki}L}{2E}$.

Let us notice that in order to obtain from (\ref{genexp5}) tree and more neutrino transition probabilities additional relations, based on the unitarity of the mixing matrix, must be used. We will
present here more simple and direct way of the derivation of the
neutrino (antineutrino) transition probabilities in which the unitarity of the mixing matrix will be fully
utilized and dependence on character of the neutrino mass spectrum will be clearly visible \cite{Bilpheno}.

It is obvious that the expression  (\ref{genexp})  can be
written in the form
\begin{equation}\label{Genexp}
  P(\nu_{\alpha}\to \nu_{\alpha'})=|\sum_{i}U_{\alpha'i}~e^{-i(E_{i}-E_{p})t}~
U^{*}_{\alpha i}|^{2},
\end{equation}
where  $p$ is an arbitrary fixed index. Further, taking into account the
unitarity of the matrix $U$, we can rewrite the transition
probability in the form
\begin{eqnarray}\label{Genexp1}
  P(\nu_{\alpha}\to \nu_{\alpha'})&=&|\delta_{\alpha'\alpha}+
\sum_{i\neq p}U_{\alpha'i}~(~e^{-i(E_{i}-E_{p})t}~-1)~
U^{*}_{\alpha i}|^{2}\nonumber\\&=&|\delta_{\alpha'\alpha}-2i\sum_{i\neq p}U_{\alpha'i}~
U^{*}_{\alpha i}e^{-i\Delta_{pi}}\sin\Delta_{pi}|^{2}
\end{eqnarray}
where
\begin{equation}\label{probability3}
\Delta _{pi}=(E_{i}-E_{p})t\simeq \frac{\Delta m^{2}_{pi}L}{4E}
\end{equation}
From (\ref{Genexp1}) we can easily obtain the following general expression for $\nu_{\alpha}\to \nu_{\alpha'}$ ($\bar\nu_{\alpha}\to \bar\nu_{\alpha'}$) transition probability
\begin{eqnarray}
P(\nua{\alpha}\to \nua{\alpha'})
=\delta_{\alpha' \alpha }
-4\sum_{i}|U_{\alpha i}|^{2}(\delta_{\alpha' \alpha } - |U_{\alpha' i}|^{2})\sin^{2}\Delta_{pi}\nonumber\\
+8~\sum_{i>k}\mathrm{Re}~U_{\alpha' i}U^{*}_{\alpha i}U^{*}_{\alpha'
k}U_{\alpha k}\cos(\Delta_{pi}-\Delta_{pk})\sin\Delta_{pi}\sin\Delta_{pk}\nonumber\\
\pm 8~\sum_{i>k}\mathrm{Im}~U_{\alpha' i}U^{*}_{\alpha i}U^{*}_{\alpha'
k}U_{\alpha k}\sin(\Delta_{pi}-\Delta_{pk})\sin\Delta_{pi}\sin\Delta_{pk}~~(i,k \neq p)
\label{Genexp4}
\end{eqnarray}

\begin{center}
{\bf Two-neutrino mixing}
\end{center}
Let us consider the simplest case of the two-neutrino mixing
\begin{equation}\label{2nu}
  \nu_{\alpha}=\sum _{i=1,2}U_{\alpha i}\nu_{iL},
\end{equation}
where the matrix $U$ has the form
\begin{eqnarray}\label{2nu1}
U =\left(
\begin{array}{cc}
\cos\theta_{12}& \sin\theta_{12}\\
-\sin\theta_{12}&\cos\theta_{12}
\end{array}
\right),
\end{eqnarray}
$\theta_{12}$ being the mixing angle. We will label neutrino masses in such a way that $m_{1}<m_{2}$ and
choose $p=1$. In this case $i=k=2$ and  we have
\begin{equation}\label{2nu2}
P(\nu_{\alpha}\to \nu_{\alpha'})=P(\bar\nu_{\alpha}\to
\bar\nu_{\alpha'})=\delta_{\alpha'\alpha}-4|U_{\alpha
2}|^{2}(\delta_{\alpha'\alpha}-|U_{\alpha'2}|^{2})\sin^{2}\Delta_{12}
.
\end{equation}
For $\alpha'\neq \alpha$ we find the following standard expression for the transition probability
\begin{eqnarray}\label{2nu3}
P(\nu_{\alpha}\to \nu_{\alpha'})&=&P(\bar\nu_{\alpha}\to
\bar\nu_{\alpha'}) =4|U_{\alpha 2}|^{2}|U_{\alpha'2}|^{2}~\sin^{2}\Delta_{12}
\nonumber\\
&=&\frac{1}{2}\sin^{2}2\theta_{12}(1-\cos\frac{\Delta
m^{2}_{12}L}{2E}).
\end{eqnarray}

For the probability of $\nu_{\alpha}$ ($\bar\nu_{\alpha}$) to survive
from (\ref{2nu2}) we find
\begin{eqnarray}\label{2nu4}
P(\nu_{\alpha}\to \nu_{\alpha})&=&P(\bar\nu_{\alpha}\to
\bar\nu_{\alpha})=1-4|U_{\alpha 2}|^{2}(1-|U_{\alpha 2}|^{2})~
\sin^{2}\Delta_{12}\nonumber\\
&=&1-\frac{1}{2}\sin^{2}2\theta_{12}(1-\cos\frac{\Delta
m^{2}_{12}L}{2E}).
\end{eqnarray}
From (\ref{2nu2}) and the unitarity relation
\begin{equation}\label{2nu5}
|U_{\alpha2}|^{2}=1-|U_{\alpha'2}|^{2}\quad (\alpha\neq \alpha').
\end{equation}
it follows that
\begin{equation}\label{2nu6}
P(\nua{\alpha}\to \nua{\alpha})=P(\nua{\alpha'}\to
\nua{\alpha'})\quad (\alpha\neq \alpha').
\end{equation}
It is obvious that this relation is valid only for the two-neutrino mixing.
\begin{center}
{\bf Three-neutrino mixing. General case}
\end{center}
In the case of the three-neutrino mixing there are two independent mass-squared differences. From analysis of the data of  neutrino oscillation experiments it follows that one  mass-squared difference is much smaller than the other one. Correspondingly, two three-neutrino  mass spectra are possible
\begin{enumerate}
  \item Normal spectrum (NS)
\begin{equation}\label{NS}
 m_{1}<m_{2}<m_{3},\quad \Delta m^{2}_{12}\ll \Delta m^{2}_{23}.
\end{equation}
\item Inverted spectrum (IS)\footnote{Notice that neutrino masses are labeled differently in the case of NS and IS. This allows to use the same notations for mixing angles in both cases.}

\begin{equation}\label{IS}
m_{3}<m_{1}<m_{2},\quad \Delta m^{2}_{12}\ll |\Delta m^{2}_{13}|.
\end{equation}
\end{enumerate}
Let us denote two independent (positive) neutrino mass-squared differences $\Delta m^{2}_{S}$ (solar) and $\Delta m^{2}_{A}$ (atmospheric). We have
\begin{equation}\label{masssquared}
\Delta
m^{2}_{12}=\Delta
m^{2}_{S},~~\Delta m^{2}_{23}=\Delta
m^{2}_{A}~~(NS)\quad \Delta
m^{2}_{12}=\Delta
m^{2}_{S},~~\Delta m^{2}_{13}=-\Delta
m^{2}_{A}~~(IS).
\end{equation}
In the case of the NS it is natural to choose $p=2$. From the general expression (\ref{Genexp4}) we have
\begin{eqnarray}
&&P^{NS}(\nua{l}\to \nua{l'})
=\delta_{l' l }
-4\sum_{i}|U_{l 1}|^{2}(\delta_{l' l} - |U_{l' 1}|^{2})\sin^{2}\Delta_{S}\nonumber\\&&-4\sum_{i}|U_{l 3}|^{2}(\delta_{l' l} - |U_{l' 3}|^{2})\sin^{2}\Delta_{A}
-8~\mathrm{Re}~U_{l' 3}U^{*}_{l 3}U^{*}_{l'
1}U_{l 1}\cos(\Delta_{A}+\Delta_{S})\sin\Delta_{A}\sin\Delta_{S}\nonumber\\
&&\mp 8~\mathrm{Im}~U_{l' 3}U^{*}_{l 3}U^{*}_{l'
1}U_{l 1}\sin(\Delta_{A}+\Delta_{S})\sin\Delta_{A}\sin\Delta_{S}
\label{Genexp5}
\end{eqnarray}
In the case of the IS we choose $p=1$. For the transition probability
we obtain the following expression
\begin{eqnarray}
&&P^{IS}(\nua{l}\to \nua{l'})
=\delta_{l' l }
-4\sum_{i}|U_{l 2}|^{2}(\delta_{l' l } - |U_{l' 2}|^{2})\sin^{2}\Delta_{S}\nonumber\\&&-4\sum_{i}|U_{l 3}|^{2}(\delta_{l' l} - |U_{l' 3}|^{2})\sin^{2}\Delta_{A}
-8~\mathrm{Re}~U_{l' 3}U^{*}_{l 3}U^{*}_{l'
2}U_{l 2}\cos(\Delta_{A}+\Delta_{S})\sin\Delta_{A}\sin\Delta_{S}\nonumber\\
&&\pm 8~\sum_{i>k}\mathrm{Im}~U_{l' 3}U^{*}_{l 3}U^{*}_{l'
2}U_{l 2}\sin(\Delta_{A}+\Delta_{S})\sin\Delta_{A}\sin\Delta_{S}
\label{Genexp6}
\end{eqnarray}
Notice that  (\ref{Genexp6}) we can obtain from (\ref{Genexp5}) if we change  $U_{l 1}\to U_{l 2}$
the sign of the last term.

From (\ref{Genexp5}) and (\ref{Genexp6}) for CP asymmetry
$A^{CP}_{l' l }=P(\nu_{l}\to \nu_{l'})-P(\bar\nu_{l}\to \bar\nu_{l'})$ we find
\begin{equation}\label{CPNH}
A^{CP}_{l' l }=-16~\mathrm{Im}~U_{l' 3}U^{*}_{l 3}U^{*}_{l'
1}U_{l 1}\sin(\Delta_{A}+\Delta_{S})\sin\Delta_{A}\sin\Delta_{S}
\end{equation}
in the case of NS and
\begin{equation}\label{CPIH}
A^{CP}_{l' l }=16~\mathrm{Im}~U_{l' 3}U^{*}_{l 3}U^{*}_{l'
2}U_{l 2}\sin(\Delta_{A}+\Delta_{S})\sin\Delta_{A}\sin\Delta_{S}
\end{equation}
in the case of IS.

In the standard parameterization the $3\times3$ PMNS \cite{BPonte57,BPonte58,MNS} mixing matrix $U$ is characterized by three mixing angles $\theta_{12},\theta_{23},\theta_{13}$ and one $CP$ phase $\delta$ and has the
form
\begin{eqnarray}
U=\left(\begin{array}{ccc}c_{13}c_{12}&c_{13}s_{12}&s_{13}e^{-i\delta}\\
-c_{23}s_{12}-s_{23}c_{12}s_{13}e^{i\delta}&
c_{23}c_{12}-s_{23}s_{12}s_{13}e^{i\delta}&c_{13}s_{23}\\
s_{23}s_{12}-c_{23}c_{12}s_{13}e^{i\delta}&
-s_{23}c_{12}-c_{23}s_{12}s_{13}e^{i\delta}&c_{13}c_{23}
\end{array}\right).
\label{unitmixU1}
\end{eqnarray}
Here $c_{12}=\cos\theta_{12}$,  $s_{12}=\sin\theta_{12}$ etc.

\begin{center}
{\bf Three-neutrino mixing. Leading approximation}
\end{center}
From analysis of the neutrino oscillation data it follows that the ratio $\frac{\Delta m^{2}_{S}}{\Delta m^{2}_{A}}$ and the value of the parameter $\sin^{2}\theta_{13}$ are small:
\begin{equation}\label{leading1}
 \frac{\Delta m^{2}_{S}}{\Delta m^{2}_{A}}\simeq 3\cdot 10^{-2},\quad
\sin^{2}\theta_{13}\simeq 2.4\cdot 10^{-2}.
\end{equation}
In atmospheric region  of the parameter $\frac{L}{E}$ ($\frac{\Delta
m^{2}_{A}L}{2E}\gtrsim 1$) in the first, leading  approximation we can neglect small contributions of
$\Delta m^{2}_{S}$ and $\sin^{2}\theta_{13}$ into neutrino transition
probabilities.

From (\ref{Genexp5}), (\ref{Genexp6})
for the probability of $\nu_{\mu}$ ($\bar\nu_{\mu}$) to survive we obtain the following expression (for
both neutrino mass spectra)
\begin{eqnarray}\label{leading1}
P(\nua{l}\to \nua{l'})
\simeq 1- 4 |U_{\mu 3}|^{2}(1-|U_{\mu 3}|^{2})\sin^{2}\Delta
m^{2}_{A}\frac{L}{4E}\nonumber\\=1-\frac{1}{2}\sin^{2}2\theta_{23}(1-\cos\Delta
m^{2}_{A}\frac{L}{2E}).
\end{eqnarray}
In the leading approximation we have $P(\nu_{\mu}\to \nu_{e})\simeq 0$ and
\begin{equation}\label{leading4}
P(\nu_{\mu}\to \nu_{\tau})\simeq 1-P(\nu_{\mu}\to \nu_{\mu})\simeq
\sin^{2}2\theta_{23}(1-\cos\Delta
m^{2}_{A}\frac{L}{2E}).
\end{equation}
Thus, in the atmospheric region predominantly two-neutrino
$\nu_{\mu}\rightleftarrows \nu_{\tau}$ oscillations take place.

Let us consider now $\bar\nu_{e}\to \bar\nu_{e}$ transition in the
reactor KamLAND region ($\frac{\Delta m^{2}_{S}L}{2E}\gtrsim 1$)   .
Neglecting contribution of $\sin^{2}\theta_{13}$ we have (for
both neutrino mass spectra)
\begin{equation}\label{leading2}
P(\bar\nu_{e}\to \bar\nu_{e})\simeq
1-\sin^{2}2\theta_{12}\sin^{2}\Delta
m^{2}_{S}\frac{L}{4E}.
\end{equation}
For appearance probabilities we find
\begin{equation}\label{leading6}
P(\bar\nu_{e}\to \bar\nu_{\mu})\simeq
\sin^{2}2\theta_{12} \cos^{2}\theta_{23}
\sin^{2}\Delta
m^{2}_{S}\frac{L}{4E}
\end{equation}
and
\begin{equation}\label{leading7}
P(\bar\nu_{e}\to \bar\nu_{\tau})\simeq
\sin^{2}2\theta_{12} \sin^{2}\theta_{23}
\sin^{2}\Delta
m^{2}_{S}\frac{L}{4E}
\end{equation}
We have
\begin{equation}\label{leading8}
  \frac{P(\bar\nu_{e}\to
\bar\nu_{\tau})}{P(\bar\nu_{e}\to \bar\nu_{\mu})}\simeq
\tan^{2}\theta_{23}\simeq 1
\end{equation}
and
\begin{equation}\label{leading9}
P(\bar\nu_{e}\to \bar\nu_{e})= 1-P(\bar\nu_{e}\to \bar\nu_{\mu})
-P(\bar\nu_{e}\to \bar\nu_{\tau})
\end{equation}
Thus, in the reactor Kamland region $\bar\nu_{e}\rightleftarrows
\bar\nu_{\mu}$ and $\bar\nu_{e}\rightleftarrows \nu_{\tau}$
oscillations take place.

The expressions (\ref{leading1}) and (\ref{leading2}) were used for analysis of the first Super-Kamiokande atmospheric data, K2K and MINOS accelerator data and data of the reactor KamLAND experiment. Now with improved accuracy of the neutrino oscillation experiments  it is more common to perform more complicated three-neutrino analysis of the data.

\begin{center}
{\bf Probability of the $\bar\nu_{e}\to \bar\nu_{e}$ transition}
\end{center}

From (\ref{Genexp5}) and  (\ref{Genexp6}) we can easily obtain exact three-neutrino expressions for $\bar\nu_{e}\to \bar\nu_{e}$
survival probabilities for both neutrino mass spectra. We have, correspondingly,
\begin{eqnarray}
&&P^{\mathrm{NS}}(\bar\nu_{e}\to \bar\nu_{e})=1- 4~|U_{e
3}|^{2}(1-|U_{e3}|^{2})~ \sin^{2}\frac{\Delta m^{2}_{A}L}{4E}
\nonumber\\
&&- 4~|U_{e 1}|^{2}(1-|U_{e 1}|^{2})~ \sin^{2}\frac{\Delta m^{2}_{S}L}{4E}
\nonumber\\
&&-8~|U_{e 3}|^{2}|U_{e
1}|^{2}~\cos\frac{(\Delta
m^{2}_{A}+\Delta m^{2}_{S})L}{4E}
\sin\frac{\Delta m^{2}_{A}L}{4E}\sin\frac{\Delta m^{2}_{S}L}{4E}.\label{3nue1}
\end{eqnarray}
and
\begin{eqnarray}
&&P^{\mathrm{IS}}(\bar\nu_{e}\to \bar\nu_{e})=1- 4~|U_{e
3}|^{2}(1-|U_{e3}|^{2})~ \sin^{2}\frac{\Delta m^{2}_{A}L}{4E}
\nonumber\\
&&- 4~|U_{e 2}|^{2}(1-|U_{e 2}|^{2})~ \sin^{2}\frac{\Delta m^{2}_{S}L}{4E}
\nonumber\\
&&-8~|U_{e 3}|^{2}|U_{e
2}|^{2}~\cos\frac{(\Delta
m^{2}_{A}+\Delta m^{2}_{S})L}{4E}
\sin\frac{\Delta m^{2}_{A}L}{4E}\sin\frac{\Delta m^{2}_{S}L}{4E}.\label{3nue2}
\end{eqnarray}
In the standard parameterization  of the PMNS mixing matrix we have
\begin{eqnarray}
&&P^{\mathrm{NS}}(\bar\nu_{e}\to \bar\nu_{e})=1-
\sin^{2}2\theta_{13} \sin^{2}\frac{\Delta m^{2}_{A}L}{2E}
\nonumber\\
&&-
(\cos^{2}\theta_{13}\sin^{2}2\theta_{12}+\sin^{2}2\theta_{13}\cos^{4}
\theta_{12}) ~ \sin^{2}\frac{\Delta m^{2}_{S}L}{2E}
\nonumber\\
&&-2\sin^{2}2\theta_{13}\cos^{2}\theta_{12} ~\cos\frac{(\Delta
m^{2}_{A}+\Delta m^{2}_{S})L}{4E} \sin\frac{\Delta
m^{2}_{A}L}{4E}\sin\frac{\Delta m^{2}_{S}L}{4E}.\label{3nue4}
\end{eqnarray}
and
\begin{eqnarray}
&&P^{\mathrm{IS}}(\bar\nu_{e}\to \bar\nu_{e})=1-
\sin^{2}2\theta_{13} \sin^{2}\frac{\Delta m^{2}_{A}L}{2E}
\nonumber\\
&&-
(\cos^{2}\theta_{13}\sin^{2}2\theta_{12}+\sin^{2}2    \theta_{13}\sin^{4}
\theta_{12}) ~ \sin^{2}\frac{\Delta m^{2}_{S}L}{2E}
\nonumber\\
&&-2\sin^{2}2\theta_{13}\sin^{2}\theta_{12} ~\cos\frac{(\Delta
m^{2}_{A}+\Delta m^{2}_{S})L}{4E} \sin\frac{\Delta
m^{2}_{A}L}{4E}\sin\frac{\Delta m^{2}_{S}L}{4E}.\label{3nue5}
\end{eqnarray}

\begin{center}
{\bf Probability of the $ \nu_{\mu}\to \nu_{e}$ ($\bar\nu_{\mu}\to \bar\nu_{e}$)
 transition }
\end{center}

From (\ref{Genexp5}) and (\ref{Genexp6}) we  obtain the following expressions for $\nua{\mu}\to \nua{e}$ vacuum transition probabilities:
\begin{eqnarray}
&&P^{\mathrm{NS}}(\nua{\mu}\to \nua{e})= 4~|U_{e 3}|^{2}|U_{\mu
3}|^{2}~ \sin^{2}\Delta_{A}
\nonumber\\
&&+4~|U_{e 1}|^{2}|U_{\mu 1}|^{2}~
\sin^{2}\Delta_{S}\nonumber\\
&&-8~\mathrm{Re}~U_{e 3}U^{*}_{\mu3}U_{e 1}^{*}U_{\mu 1}~\cos(\Delta
_{A}+\Delta _{S}) \sin\Delta
_{A}\sin\Delta _{S}\nonumber\\
&&\mp 8~\mathrm{Im}~U_{e 3}U^{*}_{\mu 3}U_{e 1}^{*}U_{\mu
1}~\sin(\Delta _{A}+\Delta _{S}) \sin\Delta _{A}\sin\Delta _{S}
.\label{3numue1}
\end{eqnarray}
and
\begin{eqnarray}
&&P^{\mathrm{IS}}(\nua{\mu}\to \nua{e})= 4~|U_{e 3}|^{2}|U_{\mu
3}|^{2}~ \sin^{2}\Delta_{A}
\nonumber\\
&&+4~|U_{e 2}|^{2}|U_{\mu 2}|^{2}~
\sin^{2}\Delta_{S}\nonumber\\
&&-8~\mathrm{Re}~U_{e 3}U^{*}_{\mu3}U_{e 2}^{*}U_{\mu 2}~\cos(\Delta
_{A}+\Delta _{S}) \sin\Delta
_{A}\sin\Delta _{S}\nonumber\\
&&\pm 8~\mathrm{Im}~U_{e 3}U^{*}_{\mu 3}U_{e 2}^{*}U_{\mu
2}~\sin(\Delta _{A}+\Delta _{S}) \sin\Delta _{A}\sin\Delta _{S}
.\label{3numue2}
\end{eqnarray}
Using the standard parameterization of the PMNS mixing
matrix in the case of NS we have
\begin{eqnarray}
&&P^{\mathrm{NS}}(\nua{\mu}\to
\nua{e})=\sin^{2}2\theta_{13}s^{2}_{23}\sin^{2}\frac{\Delta
m^{2}_{A}L}{4E}\nonumber\\&&+(\sin^{2}2\theta_{12}c^{2}_{13}c^{2}_{23}+
\sin^{2}2\theta_{13}c^{4}_{12}s^{2}_{23}+Kc^{2}_{12}\cos\delta)
\sin^{2}\frac{\Delta
m^{2}_{S}L}{4E}\nonumber\\&&+(2\sin^{2}2\theta_{13}s^{2}_{23}c^{2}_{12}+K\cos\delta)
~\cos\frac{(\Delta m^{2}_{A}+\Delta m^{2}_{S})L}{4E} \sin\frac{\Delta
m^{2}_{A}L}{4E}\sin\frac{\Delta m^{2}_{S}L}{4E}\nonumber\\
&&\mp K\sin\delta~~\sin\frac{(\Delta m^{2}_{A}+\Delta
m^{2}_{S})L}{4E} \sin\frac{\Delta m^{2}_{A}L}{4E}\sin\frac{\Delta
m^{2}_{S}L}{4E}. \label{3numue3}
\end{eqnarray}
Here
\begin{equation}\label{3numue4}
  K=\sin2\theta_{12}\sin2\theta_{13}\sin2\theta_{23}c_{13}.
\end{equation}
In the case of the inverted neutrino mass spectrum we find the
following expressions for the transition probabilities
\begin{eqnarray}
&&P^{\mathrm{IS}}(\nua{\mu}\to
\nua{e})=\sin^{2}2\theta_{13}s^{2}_{23}\sin^{2}\frac{\Delta
m^{2}_{A}L}{4E}\nonumber\\&&+(\sin^{2}2\theta_{12}c^{2}_{13}c^{2}_{23}+
\sin^{2}2\theta_{13}s^{4}_{12}s^{2}_{23}-Ks^{2}_{12}\cos\delta)
\sin^{2}\frac{\Delta
m^{2}_{S}L}{4E}\nonumber\\&&+(2\sin^{2}2\theta_{13}s^{2}_{23}s^{2}_{12}-K\cos\delta)
~\cos\frac{(\Delta m^{2}_{A}+\Delta m^{2}_{S})L}{4E} \sin\frac{\Delta
m^{2}_{A}L}{4E}\sin\frac{\Delta m^{2}_{S}L}{4E}\nonumber\\
&&\mp K\sin\delta~~\sin\frac{(\Delta m^{2}_{A}+\Delta
m^{2}_{S})L}{4E} \sin\frac{\Delta m^{2}_{A}L}{4E}\sin\frac{\Delta
m^{2}_{S}L}{4E}. \label{3numue5}
\end{eqnarray}
Formulas (\ref{3numue3}) and (\ref{3numue5}) can be used for analysis of the data of the MINOS and T2K long baseline accelerator  experiments in which matter effects can be neglected.

\section{Status of neutrino oscillations}

We will briefly discuss here the results that were obtained in recent
neutrino oscillation experiments.
\begin{center}
{\bf The SNO solar neutrino experiment}
\end{center}
The SNO experiment \cite{SNO} was performed in the Creighton mine ( Sudbury,
Canada, depth  2092 m). Solar neutrino were detected by a large
heavy-water detector (1000 tons of  D$_2$O contained in an acrylic
vessel of 12 m in diameter). The detector was equipped with 9456
photo-multipliers to detect light created by particles which are
produced in neutrino interaction.

The high-energy $^{8}B$ neutrinos were detected in the SNO
experiment. An  important feature of the SNO experiment was the
observation of solar neutrinos {\em via three different processes}.
    \begin{enumerate}
  \item The  CC process
\begin{equation}\label{SNO-CC}
\nu_e + d \to e^- + p + p ~.
\end{equation}
  \item The NC process
\begin{equation}\label{SNO-NC}
\nu_x + d \to \nu_x + p + n \quad (x=e,\mu,\tau)
\end{equation}
  \item Elastic neutrino-electron scattering (ES)
\begin{equation}\label{SNO-ES}
\nu_x + e \to \nu_x + e~.
\end{equation}
\end{enumerate}
The detection of solar neutrinos through the observation of the NC
reaction (\ref{SNO-NC}) allows one to determine the total flux of
$\nu_{e}$, $\nu_{\mu}$ and $\nu_{\tau}$ on the earth. In the SNO
experiment it was found
\begin{equation}\label{totalflux}
\Phi^{NC}_{\nu_{e,\mu,\tau}}=(5.25\pm 16 (\mathrm{stat}) {}^{+0.11}_{-0.13}(\mathrm{syst}))
\cdot 10^{6}~\mathrm{cm}^{-2}~\mathrm{s}^{-1}.
\end{equation}
The total flux of all active neutrinos on the earth must be equal to
the total flux of $\nu_{e}$ emitted by the sun (if there are no
transitions of $\nu_{e}$ into sterile neutrinos). The flux measured
by SNO is in agreement with the total flux of $\nu_{e}$ predicted by
the Standard Solar Model:
\begin{equation}\label{totalflux1}
\Phi^{SSM}_{\nu_{e}}=(4.85\pm 0.58)\cdot 10^{6}~\mathrm{cm}^{-2}~\mathrm{s}^{-1}.
\end{equation}
The detection of the solar neutrinos via the reaction (\ref{SNO-CC})
allows one to determine the total flux of $\nu_{e}$ on the earth. It
was found in the SNO experiments that the total flux of $\nu_{e}$ was
about three times smaller than the total flux of all active
neutrinos.

From the ratio of the fluxes of $\nu_{e}$ and $\nu_{e}$, $\nu_{\mu}$
and $\nu_{\tau}$ the $\nu_{e}$ survival probability can be
determined. It was shown in the SNO experiment that in the
high-energy $^{8}B$ region  the $\nu_{e}$ survival probability did
not depend on the neutrino energy and was equal to
\begin{equation}\label{ratCCNC}
    \frac{\Phi^{CC}_{\nu_{e}}}{\Phi^{NC}_{\nu_{e,\mu,\tau}}}=P(\nu_{e}\to \nu_{e})=
0.317\pm 0.016 \pm 0.009.
\end{equation}
Thus, it was proved in a direct, model independent way that solar
$\nu_{e}$ on the way to the earth are transferred into $\nu_{\mu}$
and $\nu_{\tau}$.

From the three-neutrino analysis of the results of the SNO and other solar neutrino experiments (Homestake \cite{Davis}, GALLEX-GNO \cite{Gallex}, SAGE \cite{Sage}, BOREXINO \cite{Borexino})
 and the reactor KamLAND experiment \cite{Kamland}  for the neutrino oscillation parameters it was found
\begin{equation}\label{parametersSNOKL}
\Delta m^{2}_{S}=(7.41^{+0.21}_{-0.19})\cdot 10^{-5}~\mathrm{eV}^2,~~\tan^{2} \theta_{12}=0.427^{+0.033}_{-0.029},~~
\sin^{2}\theta_{13}=(2.5^{+1.8}_{-1.5})\cdot 10^{-2}.
\end{equation}

\begin{center}
{\bf The KamLAND reactor neutrino experiment}
\end{center}
The KamLAND detector \cite{Kamland} is located in the Kamioka mine (Japan) at a
depth of about 1 km. The neutrino target is a 1 kiloton liquid
scintillator which is  contained in a 13 m-diameter transparent nylon
balloon suspended in 1800 $m^{3}$ non-scintillating buffer oil. The
balloon and buffer oil are contained in an 18 m-diameter
stainless-steel vessel. On the inner surface of the vessel 1879
photomultipliers are mounted.

In the KamLAND experiment $\bar\nu_{e}$ from 55
reactors  at  distances of $175\pm 35$ km
from the Kamioka mine are detected.

Reactor $\bar\nu_{e}$'s are detected
in the KamLAND experiment through the observation of the process
\begin{equation}\label{inversebeta}
\bar\nu _{e}+p \to e^{+}+n.
\end{equation}
The signature of the neutrino event
is a coincidence between two $\gamma$-quanta produced in
the annihilation of a positron (prompt signal) and
a $\simeq 2.2$ MeV $\gamma$-quantum produced by a neutron capture
in the process $n+p\to d+\gamma$ (delayed
signal).

The average energy of the reactor antineutrinos is 3.6 MeV.
For such energies, distances of about 100~km are appropriate
 to study neutrino oscillations driven by the solar
neutrino mass-squared difference $\Delta m_{S}^{2}$.

From March 2002 to May 2007 in the KamLAND experiment 1609 neutrino
events were observed. The expected number of neutrino events (if
there are no neutrino oscillations) is $2179 \pm 89 $. Thus, it was
proved that $\bar\nu _{e}$'s disappeared on the way from the reactors
to the detector.

As the $\bar\nu _{e}$ survival probability depends on the neutrino
energy, we must expect that the detected spectrum of $\bar\nu _{e}$
is different from  the reactor antineutrino spectrum. In fact, in the
KamLAND experiment a significant distortion of the initial
antineutrino spectrum was observed.

The data of the experiment are well described if we assume that
two-neutrino oscillations take place. For the neutrino oscillation
parameters it was found
\begin{equation}\label{KLparameters}
 \Delta m_{12}^{2}=(7.66~ {}_{-0.22}^{+0.20})\cdot 10^{-5}~\mathrm{eV}^{2},\quad \tan^{2}2\theta_{12}=
0.52~{}^{+0.16}_{-0.10} .
\end{equation}
 From the three-neutrino analysis of all solar neutrino data and
the data of the KamLand reactor experiment for the neutrino
oscillation parameters the following values were obtained:
\begin{equation}\label{1solkam}
\Delta m_{12}^{S}=(7.50 {}^{+0.19}_{-0.20})\cdot 10^{-5}~\mathrm{eV}^{2},
\quad \tan^{2}\theta_{12}=
0.452{}^{+0.035}_{-0.033},\quad \sin^{2}\theta_{13}=0.020\pm 0.016.
\end{equation}

\begin{center}
{\bf  Super-Kamiokande atmospheric neutrino experiment}
\end{center}

In the Super-Kamiokande atmospheric neutrino experiment\cite{SK} the first
model-independent evidence in favor of neutrino oscillations was
obtained (1998). The Super-Kamiokande detector is located in the
same Kamioka mine as the KamLAND detector. It consists of two
optically separated water-Cherenkov cylindrical detectors with a
total mass of 50 kilotons of water. The inner detector with 11146
photomultipliers  has a radius of 16.9 m and a height of 36.2 m. The
outer detector is a veto detector. It allows to reject cosmic ray
muons. The fiducial mass of the detector is 22.5 kilotons.

In the Super-Kamiokande  experiment atmospheric neutrinos
in a wide range of energies from about 100 MeV to about 10 TeV are detected .
Atmospheric $\nu_{\mu}$ ($\bar\nu_{\mu}$) and $\nu_{e}$ ($\bar\nu_{e}$)
are detected through the observation of
$\mu^{-}$ ($\mu^{+}$) and $e^{-}$ ($e^{+}$) produced in the processes
\begin{equation}\label{SKatm}
 \nu_{\mu}(\bar \nu_{\mu}) +N\to \mu^{-}(\mu^{+}) +X, \quad
\nu_{e}(\bar \nu_{e}) +N\to e^{-}(e^{+}) +X.
\end{equation}
For the study of neutrino oscillations it is important to
distinguish electrons and muons produced in the processes (\ref{SKatm}).
In the Super-Kamiokande experiment leptons are observed through the
detection of the Cherenkov radiation. The shapes of the Cherenkov rings
of electrons and muons are completely different (in the case of
electrons  the Cherenkov rings exhibit a more diffuse light  than in
the muon case). The probability of a misidentification of electrons
and muons is below  2\%.

A model-independent evidence of neutrino oscillations was
obtained by the Super-Kamiokande Collaboration through the investigation
of the zenith-angle dependence of the  electron and muon
events. The zenith angle $\theta$ is determined in such a way that
neutrinos going vertically downward have $\theta = 0$ and neutrinos
coming vertically upward through the earth have $\theta = \pi$.
 At neutrino energies $E \gtrsim 1$ GeV the fluxes
of muon and electron neutrinos are symmetric under the change
$\theta \to \pi -\theta$. Thus, if there are no neutrino
oscillations in this energy region the numbers of electron and muon events
must satisfy the relation
\begin{equation}\label{azsym}
    N_{l}(\cos\theta)=N_{l}(-\cos\theta)\quad l=e,\mu.
\end{equation}
In the Super-Kamikande experiment a large distortion of this relation
for high energy muon events was established (a significant deficit of
upward-going muons was observed). The number of electron events
satisfies the relation (\ref{azsym}).

This result can naturally  be explained by the disappearance of muon
neutrinos due to neutrino oscillations.  The probability for
$\nu_{\mu}$ to survive depends on the distance between the neutrino
source and the neutrino detector. Downward going neutrinos ($\theta
\simeq 0$) pass a distance of about 20 km. On the other side upward
going neutrinos ($\theta \simeq \pi$) pass a distance of about 13000
km (earth diameter). The measurement of the dependence of the numbers
of the electron and muon events on the zenith angle $\theta$ allows
one to span the whole region of distances from about 20 km to about
13000 km.

From the data of the Super-Kamiokande experiment
for high-energy electron events  was found
\begin{equation}\label{updown1}
\left(\frac{U}{D}\right)_{e}=0.961{}^{+0.086}_{-0.079}\pm 0.016.
\end{equation}
For high-energy muon  events  the value
\begin{equation}\label{updown2}
\left(\frac{U}{D}\right)_{\mu}=0.551{}^{+0.035}_{-0.033}\pm 0.004.
\end{equation}
was obtained.  Here $U$ is the total number of upward going leptons
($-1< \cos\theta<-0.2$ ) and $D$ is the total number of downward
going leptons ($0.2< \cos\theta< 1$).

The data of the Super-Kamiokande atmospheric neutrino experiment are
well described if we assume that $\nu_{\mu}$'s disappear mainly due
to $\nu_{\mu}\rightleftarrows  \nu_{\tau} $ oscillations. From the
three-neutrino analysis of the data for neutrino oscillation
parameters in the case  of normal (inverted) neutrino mass spectrum
it was found
\begin{eqnarray}\label{SKanalysis}
1.9 ~(1.7)\cdot 10^{-3}~\mathrm{eV}^{2} \leq\Delta m_{A}^{2} \leq
2.6~ (2.7)\cdot 10^{-3}~\mathrm{eV}^{2},\nonumber\\
0.407 \leq \sin^2 \theta_{23} \leq 0.583,\quad
\sin^2 \theta_{13} < 0.04 (0.09).
\end{eqnarray}
The result of the Super-Kamiokande atmospheric neutrino experiment was
fully confirmed by

\begin{center}
{\bf The long-baseline MINOS accelerator neutrino experiments}

\end{center}
The first accelerator long-baseline neutrino experiment was K2K \cite{K2K}.
In the next MINOS experiment \cite{Minos} muon neutrinos produced at the Fermilab Main
Injector facility are detected. The MINOS data were obtained  with
neutrinos  with energies in the range $1\leq E\leq 5$ GeV.

There are two identical neutrino detectors in the  experiment. The
near detector with a mass of 1 kiloton is at a distance of about 1 km
from the target and about 100~ m underground. The far detector with a
mass of 5.4 kilotons is in the Sudan mine at a distance of 735 km
from the target (about 700 m underground).

Muon neutrinos (antineutrinos) are detected in the  experiment via the
observation of the process
\begin{equation}\label{Minos}
\nu_{\mu}(\bar\nu_{\mu}) +\rm{Fe}\to \mu^{-}(\mu^{+})+X
\end{equation}
The neutrino energy is given  by the sum of the muon energy
and the energy of the hadronic shower.

In the near detector the initial neutrino spectrum  is measured.
This measurement allows
to predict the expected spectrum of the muon neutrinos in the far
detector in the case if there were no neutrino oscillations.
A strong distortion of the
spectrum of $\nu_{\mu}(\bar\nu_{\mu})$
in the far detector was observed in MINOS experiment.

From the two-neutrino analysis of the $\nu_{\mu}$ data for
the neutrino oscillations parameters the following values were obtained
\begin{equation}\label{Minos1}
\Delta m_{A}^{2}=(2.32~{}_{-0.08}^{+0.12})\cdot 10^{-3}~\mathrm{eV}^{2},\quad \sin^{2}2\theta_{23}>0.90.
\end{equation}
From three-neutrino analysis of all MINOS data in the case of the normal spectrum it was found
\begin{equation}\label{Minos2}
\Delta m_{A}^{2}=(2.28-2.46)\cdot 10^{-3}~\mathrm{eV}^{2}~(68\% CL),\quad \sin^{2}\theta_{23}=(0.35-0.65)~(90\% CL).
\end{equation}
In the case of the inverted spectrum it was obtained
\begin{equation}\label{Minos3}
\Delta m_{A}^{2}=(2.32-2.53)\cdot 10^{-3}~\mathrm{eV}^{2}~(68\% CL),\quad \sin^{2}\theta_{23}=(0.34-0.67)~(90\% CL).
\end{equation}

\begin{center}

{\bf Measurements of the angle $\theta_{13}$ in reactor experiments}
\end{center}

The value of the mixing angle $\theta_{13}$ is extremely important
for the future of neutrino physics. If this angle is not equal to
zero (and relatively large) in this case it will be possible to
observe such a fundamental effect of the three-neutrino mixing as
$CP$ violation in the lepton sector. Another problem, the solution of
which requires nonzero $\theta_{13}$, is the problem of the neutrino
mass spectrum.

During many years only an upper bound on the parameter
$\sin^{2}\theta_{13}$ existed. This bound was obtained from the
analysis of the data of the reactor CHOOZ  experiment \cite{Chooz}.

From the data of the CHOOZ experiment  the following upper bound
\begin{equation}\label{Chooz2}
\sin^{2}2\,\theta_{13} \leq 0.16.
\end{equation}
was obtained.

The Double Chooz experiment \cite{Doublechooz} is performed with one antineutrino detector exposed to two reactors.
The Double Chooz collaboration published  first indication in favor
of reactor $\bar\nu_{e}$'s disappearence. For the
ratio of the observed and predicted $\bar\nu_{e}$ events the value
$0.944 \pm 0.016 \pm 0.040$ was found. From the recent background model independent analysis of
the data it was found $\sin^{2}2\theta_{13}=0.102\pm 0.028 (\mathrm{stat})\pm 0.033 (\mathrm{syst})$

In the Daya Bay experiment \cite{Dayabay} antineutrinos
from six reactors (the thermal power of each reactor is 2.9 GW) were
detected by six identical  Gd-loaded 20 ton liquid scintillator detectors located in two near
(flux-weighted distances 512 m and 561 m) and one far (1579 m) underground halls.
Reactor $\bar\nu_{e}$'s are detected
via observation of the standard reaction
\begin{equation}\label{Dayabay}
\bar\nu_e + p \to e^+ + n.
\end{equation}
During 217 days 41589 (203809 and 92912) antineutrino candidate-events were observed  in the detectors of the far (near) halls . From combined analysis of $\bar\nu_{e}$ rates and energy spectra measured in all six detectors it was found
\begin{equation}\label{Dayabay1}
\sin^{2}2\theta_{13}=0.090^{+0.008}_{-0.009} ,\quad \Delta m^{2}_{ee}=(2.59 ^{+0.19}_{-0.20})\cdot 10^{-3}~\mathrm{eV}^{2},
\end{equation}
where the parameter $\Delta m^{2}_{ee}$ is determined by the relation
\begin{equation}\label{Dayabay1}
\sin^{2}\Delta_{ee}= \cos^{2}\theta_{12} \sin^{2}\Delta_{13} + \sin^{2}\theta_{12} \sin^{2}\Delta_{23}, \quad
\Delta _{ik}= \frac{\Delta m^{2}_{ik}L}{4E}.
\end{equation}

In the reactor RENO experiment \cite{Reno} $\bar\nu_{e}$'s from six
reactors with total thermal power  16.5 GW were detected by  near and
far detectors (Gd-loaded 16-ton liquid scintillators). The detectors are located at 294 m and 1383 m
from the center of the reactor array. During 222 days in the far and
near detectors 30211 and 279787 candidate-events were observed,
respectively. For the ratio of the observed and predicted
antineutrino events in the far detector it was found the value
\begin{equation}\label{Reno1}
  R=0.929\pm 0.006 \pm 0.007.
\end{equation}
From rate-only analysis of the data it was found

\begin{equation}\label{Reno2}
 \sin^{2}2\theta_{13}=0.100 \pm 0.010 \pm 0.015.
\end{equation}

\begin{center}

{\bf Accelerator T2K experiment}
\end{center}
The value of the parameter $\theta_{13}$ was measured in the accelerator
 long baseline  T2K neutrino experiment
\cite{T2K}. In this experiment muon neutrinos produced at the J-PARC
accelerator  in Japan are detected at a distance of 295 km in the
water-Cherenkov Super-Kamiokande detector. The T2K experiment is the
first off-axis neutrino experiment: the angle between the direction
to the detector and the flight direction of the parent $\pi^{+}$'s is
equal to
 $2^\circ$. This allows one to obtain a narrow-band neutrino beam
with a maximal intensity at the energy  $E\simeq 0.6 $ GeV  which corresponds at
the distance of $L=295$ km to the first oscillation maximum
($E_{0}=\frac{2.54}{\pi}\Delta m_{A}^{2}L$).

At a distance of about 280 m from the target there are several
near detectors which are
used for the measurement of the neutrino spectrum and flux  and  for
the measurement of cross sections of different CC and NC processes.

The initial beam (from decays of pions and kaons) is a beam of
$\nu_{\mu}$'s with a small (about 0.4 \%) admixture of $\nu_{e}$'s.
The search for electrons in the Super-Kamiokande detector due to
$\nu_{\mu}\to \nu_{e}$ transitions was performed and  28 $\nu_{e}$
events were observed. The expected background is equal to $4.92\pm
0.55$ events. In the case of the three-neutrino mixing $\nu_{\mu}\to \nu_{e}$ transition probability depends on  6 neutrino oscillation parameters. Assuming that $\sin^{2}\,\theta_{12}=0.306$, $\Delta m^{2}_{S}=7.6\cdot 10^{-5}~\mathrm{eV}^{2}$,
$\sin^{2}\,\theta_{23}=0.5$, $\Delta m^{2}_{A}=2.4\cdot 10^{-3}~\mathrm{eV}^{2}$, $\delta=0$
from the analysis of the T2K data for the normal neutrino mass
spectrum it was found:
\begin{equation}\label{t2k}
  \sin^{2}2\,\theta_{13}=0.140^{+0.038}_{-0.032}.
\end{equation}
For the inverted neutrino mass spectrum it was obtained
\begin{equation}\label{t2k1}
 \sin^{2}2\,\theta_{13}=0.170^{+0.045}_{-0.037}.
\end{equation}

Recently T2K Collaboration published their first results on the measurement of the parameters $\sin^{2}2\,\theta_{23}$ and $\Delta m^{2}_{A}$. From the analysis of the $\nu_{\mu}$ disappearance data it was found
\begin{equation}\label{T2K}
\sin^{2}\,\theta_{23}=0.514^{+0.055}_{-0.056},\quad \Delta m^{2}_{A}=(2.51\pm 0.10)\cdot 10^{-3}~\mathrm{eV}^{2},~~{\bf NS}
\end{equation}
and
\begin{equation}\label{T2K1}
\sin^{2}\,\theta_{23}=0.511^{+0.055}_{-0.056},\quad \Delta m^{2}_{A}=(2.48\pm 0.10)\cdot 10^{-3}~\mathrm{eV}^{2},~~{\bf IS}
\end{equation}

\section{On the problem of the neutrino masses }
It was established by the neutrino oscillation  experiments that neutrino masses are different from zero and flavor neutrino states are mixtures of states of neutrinos with definite masses. However, the origin of neutrino masses and neutrino mixing is still an open problem. In this section we will briefly discuss this problem.

We will start with the following remark. Left-handed and right-handed quark and charged lepton fields and left-handed neutrino fields $\nu_{lL}$ ($l=e, \mu, \tau$) are SM fields (left-handed fields are components of $SU(2)$ doublets and right-handed fields are singlets). What about right-handed neutrino fields $\nu_{lR}$? If $\nu_{lR}$ are also SM fields (singlets) in this case  neutrino masses (like quark and lepton masses) can be generated by the standard Higgs mechanism via Yukawa interaction
\begin{equation}\label{Yukawa}
\mathcal{L}^{Y}_{I}=-\sqrt{2}\sum_{l'l}\overline  L_{l'L}Y_{l'l}\nu_{lR}\tilde{H }+\mathrm{h.c.}.
\end{equation}
Here
\begin{eqnarray}\label{doublets}
L_{lL}
=
\left(
\begin{array}{c}
\nu_{lL}
\\
l_L
\end{array}
\right)
\qquad
H
=
\left(
\begin{array}{c}
H^{(+)}
\\
H^{(0)}
\end{array}
\right)
\end{eqnarray}
are the lepton and Higgs doublets, $\tilde{H}=i\tau_{2}H^{*}$, $Y_{l'l}$ are
dimensionless Yukawa constants.

After spontaneous breaking of the electroweak symmetry we have
\begin{eqnarray}\label{break}
\tilde{H}
=
\left(
\begin{array}{c}
\frac{v+H}{\sqrt{2}}
\\
0
\end{array}
\right),
\end{eqnarray}
where $H$ is the Higgs field. From (\ref{Yukawa}) and (\ref{break}) we obtain the Dirac mass term
\begin{equation}\label{Dmass}
 \mathcal{L}^{\mathrm{D}}=-\sum_{l',l}{\bar\nu_{l'L}}\,M_{l'l}^{\mathrm{D}} \nu_{lR}
+\mathrm{h.c.}.
\end{equation}
Here
\begin{equation}\label{Dmass1}
M_{l'l}^{\mathrm{D}}=v~Y_{l'l},
\end{equation}
where $v=(\sqrt{2}G_{F})^{-1/2}\simeq 246$ GeV is the vacuum expectation value of the Higgs field.

After the standard diagonalization of the matrix $Y$ ( $Y=U~y~V^{\dag}$, where $U$ and $V$ are unitary matrices and $y$ is a diagonal matrix)
for the Dirac neutrino masses we have
\begin{equation}\label{Dmass2}
    m_{i}=v~y_{i}.
\end{equation}
 Let us assume the hierarchy of neutrino masses $m_{1}\ll m_{2}\ll m_{3}$. In this case for the largest neutrino mass we have the value $m_{3}\simeq\sqrt{\Delta m_{A}^{2}}\simeq 5\cdot 10^{-2}$ eV. For the corresponding Yukawa coupling we find
\begin{equation}\label{Yukawa1}
y_{3}\simeq 2 \cdot 10^{-13}.
\end{equation}
 and $y_{1,2}\ll y_{3}$. The Yukawa couplings of quarks and lepton of the  third generation are in the range
($ 1- 10^{-2}$). Thus, neutrino Yukawa coupling is eleven (!) or more orders of magnitude smaller that Yukawa couplings
of quarks and the lepton.  Such extremely small values of the Yukawa couplings of neutrinos are commonly considered as a strong argument against SM origin of the neutrino masses.

Several beyond the SM mechanisms of neutrino mass generation were proposed. Apparently, {\em the most plausible and the most viable is the seesaw mechanism}\cite{seesaw}. We will consider here the  approach which is based on the effective Lagrangian formalism \cite{Weinberg}. Effective Lagrangians describe effects of a beyond the SM physics in the SM observables. They are invariant under electroweak $SU(2)\times U(1)$ transformations and are nonrenormalizable.

In order to generate neutrino masses we need
 to build an effective Lagrangian which is quadratic in the lepton fields. The term
$\bar L_{lL}\tilde{H }$ is $SU(2)\times U(1)$ invariant and has dimension $M^{5/2}$. We can build the effective Lagrangian in the same way as the Majorana mass term was built. Taking into account that Lagrangians must have dimension four we have
\begin{equation}\label{effective}
 \mathcal{L}_{I}^{\mathrm{eff}}=-\frac{1}{\Lambda}~\sum_{l'l}\bar L_{l'L}\tilde{H }\bar Y_{l'l}\tilde{H }^{T} L^{c}_{lL}+\mathrm{h.c.},
\end{equation}
 where the parameter $\Lambda$ has the dimension of mass. It characterizes the scale at which the lepton number $L$ is violated. We assume that $\Lambda\gg v$.\footnote{Let us notice that for the dimensional arguments we used here it was important that Higgs is not composite particle and exist Higgs field having dimension $M$. Recent discovery of the Higgs boson at CERN \cite{Higgs} confirm this assumption.}

After spontaneous breaking of the electroweak symmetry  from (\ref{break}) and (\ref{effective}) we obtain the Majorana mass term
(\ref{Mjmst}) in which the matrix $ M^{\mathrm{M}}$ is given by the relation
\begin{equation}\label{Mjmass2}
M^{\mathrm{M}}=\frac{v^{2}}{\Lambda}~\bar Y.
\end{equation}
For the neutrino mass we have
\begin{equation}\label{numass}
    m_{i}=\frac{v^{2}}{\Lambda}~\bar y_{i}.
\end{equation}
Thus, if Majorana neutrino masses are generated by a beyond the SM mechanism the value of the neutrino mass
 $m_{i}$ is a product of a "typical fermion mass" $v~\bar y_{i}$ and a suppression factor which is given by the ratio of the electroweak scale $v$ and a scale $\Lambda$ of a new lepton-number violating physics.

In order to estimate $\Lambda$ we assume hierarchy of neutrino masses. In this case $m_{3}\simeq 5\cdot 10^{-2}$ eV. Assuming also that the parameter $\bar y_{3}$ is of the order of one we find $\Lambda\simeq 10^{15}$ GeV. Hence, smallness of the Majorana neutrino masses could mean that $L$ is violated at a large scale.

Violation of the lepton number can be connected with the existence of heavy Majorana leptons which interact with the SM particles \cite{Weinberg80}. Let us assume that heavy Majorana leptons $N_{i}$ ($i=1,...n$), singlets of $SU(2)\times U(1)$ group, have the following Yukawa lepton-number violating interaction
\begin{equation}\label{HeavyMj}
 \mathcal{L}_{I}^{Y}=-\sqrt{2}\sum _{l,i}y_{li} \bar L_{lL}\tilde{H }N_{iR}+\mathrm{h.c.}.
\end{equation}
Here $N_{i}=N^{c}_{i}$ is the field of a Majorana lepton with the mass $M_{i}\gg v$ and $y_{li}$ are dimensionless Yukawa coupling constants.

In the second order of the perturbation theory at the electroweak energies  the interaction (\ref{HeavyMj}) generates the effective Lagrangian (\ref{effective}) in which the constant $\frac{\bar Y_{l'l}}{\Lambda}$ is given by the relation
\begin{equation}\label{HeavyMj1}
 \frac{\bar Y_{l'l}}{\Lambda}=\sum_{i}y_{l'i}\frac{1}{M_{i}} y_{li}.
\end{equation}
From this relation it follows that the scale of a new lepton-number violating physics is determined by masses of heavy Majorana leptons.

Summarizing, if neutrino masses are generated by the interaction of the lepton and Higgs doublets with heavy Majorana leptons in this case
\begin{enumerate}
  \item Neutrinos with definite masses are Majorana paricles.
  \item Neutrino masses are much smaller then   masses of charged leptons and quarks.
They are of the order of "standard fermion mass" multiplied by a suppression factor which is given by the ratio of  Higgs vacuum expectation value $v$ and a mass of heavy Majorana lepton.
\item The number of the massive neutrinos is equal to the number of lepton generations (three)\footnote{ In the scheme of neutrino masses and mixing based on the effective Lagrangian approach there are no sterile neutrinos. As it is well known, at present exist different indications in favor of the transition of flavor neutrinos into sterile states (see \cite{sterile}). Numerous experiments, now at preparation, which aim to check existence of the sterile neutrinos will confirm or disprove the scenario we have considered here.}
\end{enumerate}
From the point of view of the approach, we have considered  here, observation of the neutrinoless double $\beta$-decay
of some even-even nuclei could mean existence of heavy Majorana leptons which induce  Majorana neutrino masses. The scale of masses of heavy leptons
is determined by the parameter $\Lambda\simeq 10^{15}$ GeV. Such heavy particles can not be produced in laboratory. However, as it is well known (see, for example, \cite{Nir}) CP-violating decays of such particles in the early Universe could be the origin of the barion asymmetry of the Universe.

In conclusion let us notice that the seesaw mechanism based on the interaction (\ref{HeavyMj}) is called type I seesaw. The effective Lagrangian (\ref{effective}) can also be generated by the Lagrangian of interaction of lepton and Higgs doublets with a heavy triplet lepton (type III seesaw) and by the Lagrangian of interaction of lepton doublets and Higgs doublets with heavy triplet scalar boson (type II seesaw).

\section{Conclusion}

There are two unique properties of neutrinos which determine their
importance and their problems:
\begin{enumerate}
  \item Neutrinos have only weak interaction.
  \item Neutrinos have very small masses.
\end{enumerate}
Since neutrinos have only weak interaction, cross sections of
interaction of neutrinos with nucleons are extremely small. This
means that it is necessary to develop special methods of neutrino
detection (large detectors which often are located in underground
laboratories etc). However,
when methods of neutrino detection were developed,  neutrinos became
a unique instrument in the study of the sun (solar neutrino
experiments allow us to obtain information about the central
invisible region of the sun in which solar energy is produced in
thermonuclear reactions), in the investigation of a mechanism of the
Supernova explosion\footnote{On February 23, 1987 for the first time
antineutrinos from Supernova SN1987A in the Large Magellanic Cloud
were detected by the Kamiokande, IMB and Baksan detectors} (99\% of the
energy produced in a Supernova explosion is emitted in neutrinos), in
establishing the  quark structure of a nucleon (through the study of
the deep inelastic processes $\nu_{\mu}(\bar\nu_{\mu})+N\to
\mu^{-}(\mu^{+}+X$), etc.

In the fifties the majority of physicists believed that the neutrino
was a massless particle. This was an important, constructive
assumption. The theory of the two-component neutrino, which was based
on this assumption, inspired the creation of the phenomenological
$V-A$ theory and later became part of the Standard Model of the
electroweak interaction.

Neutrino masses are very small and it is very difficult to observe
{\em effects of neutrino masses in the $\beta$-decay and in other
weak processes.} However, small neutrino masses and, correspondingly,
mass-squared differences make possible the production (and detection)
of the {\em coherent flavor neutrino states} and quantum-mechanical periodical transitions between
different flavor neutrino neutrinos (neutrino oscillations). The
observation of neutrino oscillations at large (macroscopic) distances
allowed to resolve small neutrino mass-squared differences.

 {\bf The discovery of neutrino oscillations signifies
a new era in neutrino physics,} the era of investigation of neutrino
properties. From the analysis of the existing neutrino oscillation
data two mass-squared differences $\Delta m^{2}_{A}$ and $\Delta
m^{2}_{S}$ and two mixing parameters $\sin^2\theta_{23}$ and
$\tan^{2}\theta_{12}$ are determined with accuracies in the range
(3-12)\%. The results of the  measurements of the parameter
$\sin^{2}2\,\theta_{13}$ was recently announced by the T2K, Double Chooz, Daya Bay and
RENO  collaborations.

One of the most urgent problems  which will be addressed in the next
neutrino oscillation experiments are
\begin{enumerate}
  \item {\bf CP violation  in the lepton sector.}
  \item {\bf The character of the neutrino mass spectrum (normal or
inverted?).}
\end{enumerate}

The "large" value of  the angle $\theta_{13}$ will open the way for the investigation of
these problems in the near future.

One of the most important problems of the physics of massive and
mixed neutrinos is the problem of the nature of neutrinos with
definite masses $\nu_{i}$.

{\bf Are neutrinos with define masses Dirac particles possessing
conserved lepton number or truly neutral Majorana particles?}

The answer to this fundamental question can be obtained in
experiments on the search for neutrinoless double $\beta$-decay
($0\nu\beta\beta$-decay) of some even-even nuclei (see \cite{BilGiunti})
\begin{equation}\label{bb}
(A,Z)\to (A,Z+2)+e^{-}+e^{-}.
\end{equation}
This process is allowed only if the total lepton number is violated.
If massive neutrinos are Majorana particles, $0\nu\beta\beta$-decay
 is the second order  in the Fermi constant $G_{F}$ process with the
exchange of the virtual neutrinos between neutron-proton-electron
vertices. The matrix element of the process is proportional to the
effective Majorana mass
\begin{equation}\label{bb1}
m_{\beta\beta}=\sum_{i}U^{2}_{ei}m_{i}.
\end{equation}
Many experiments on the search for the $0\nu\beta\beta$-decay of
different  nuclei were performed. No evidence in favor of the process
was obtained. Future experiments on the search for the $0\nu\beta\beta$-decay
will be sensitive to the value
$$|m_{\beta\beta}|\simeq \rm(a~few)~10^{-2}~\rm{eV}$$
and can probe the Majorana nature of $\nu_{i}$ in the case of
the inverted hierarchy of the neutrino masses
\begin{equation}\label{Inhierar}
m_{3}\ll m_{1}<m_{2}.
\end{equation}
Another fundamental question of the neutrino physics is

{\bf What is the value of the neutrino mass?}

From the data of neutrino oscillation experiments only the
mass-squared differences can be determined. The absolute value of the
neutrino mass $m_{\beta}$ can be inferred from the investigation of
$\beta$-spectra. From the data of the  Mainz \cite{Mainz} and Troitsk
\cite{Troitsk} tritium experiments the following bounds was obtained, respectively
$$m_{\beta}< 2.3 ~\rm{eV},\quad m_{\beta}< 2.05 ~\rm{eV}$$
where $ m_{\beta}=\sqrt{\sum_{i}|U_{e1}|^{2}m^{2}_{i}}$ is the
"average" neutrino mass. The future tritium experiment KATRIN
\cite{Katrin} will be sensitive to
$$m_{\beta}< 0.2 ~\rm{eV}$$
Precision modern cosmology became an important source of information
about absolute values of neutrino masses. Different cosmological
observables (Large Scale Structure of the Universe, Gravitational
Lensing of Galaxies, Primordial Cosmic Microwave Background, etc.)
are sensitive to the sum of the neutrino masses $\sum_{i}m_{i}$. From
the existing data the following bounds were obtained \cite{Hannestad}
\begin{equation}\label{cosmo}
\sum_{i}m_{i}<(0.2-1.3)~\mathrm{eV}.
\end{equation}
It is expected that future cosmological observables will be sensitive
to the sum of neutrino masses in the range \cite{Abazajian}
\begin{equation}\label{cosmo1}
\sum_{i}m_{i}\simeq (0.05-0.6)~\mathrm{eV}.
\end{equation}
These future measurements (\ref{Inhierar}), apparently,  will probe
the inverted neutrino mass hierarchy ($\sum_{i}m_{i}\simeq 0.1$ eV)
and even the normal neutrino mass hierarchy.

The next question which needs to be answered

{\bf How many neutrinos with definite masses exist in nature?}

As we have discussed earlier the number
of massive light neutrinos can be more than three. In this case
flavor neutrinos could oscillate into sterile states, which
do not have the standard weak interaction.

For many years there existed an indication in favor of more than
three light neutrinos with definite masses obtained in a
short-baseline LSND experiment \cite{LSND}. In this experiment the
$\bar\nu_{\mu}\to \bar\nu_{e}$ transition driven by $\Delta
m^{2}\simeq 1~\mathrm{eV}^{2}$, which is much larger than the
atmospheric mass-squared difference, was observed. Some indications
in favor of more than three massive neutrinos were also obtained in
the MiniBooNE and reactor experiments (see \cite{Miniboone}). New
short-baseline accelerator and reactor experiments are urgently
needed. Such experiments are now at preparation. There are other
questions connected with neutrinos which are now being actively
discussed in the literature (neutrino magnetic moments, nonstandard
interaction of neutrinos, etc.).

An explanation of small neutrino masses requires a new, beyond the
Standard Model (Higgs) mechanism of neutrino mass generation. Several mechanisms of neutrino mass generation were
proposed. Apparently, the most viable mechanism
is the seesaw mechanism of the neutrino mass generation
\cite{seesaw}.

The seesaw mechanism is based on the assumption that the total lepton
number $L$ is violated at  a large scale $\Lambda\simeq 10^{15}$ GeV. If
this mechanism is realized, in this case neutrino masses are given in
form of  products of electroweak masses and a very small factor
$\frac{v}{\Lambda}$, which is the ratio of the electroweak scale $v\simeq
250$ GeV and a new scale $\Lambda$ which characterizes the violation of
$L$.

In order to reveal the true nature of neutrino masses and mixing many
new investigations must be performed. There are no doubts that new surprises, discoveries (and,
possibly, Nobel Prizes) are ahead.

This work is supported by the Alexander von Humboldt Stiftung, Bonn, Germany (contract Nr. 3.3-3-RUS/1002388),
and by RFBR Grant N 13-02-01442.

\end{document}